\documentclass[a4paper,12pt]{article}

\usepackage[utf8]{inputenc} 
\usepackage{hyperref}       
\usepackage{url}            
\usepackage{booktabs}       
\usepackage{amsfonts}       
\usepackage{nicefrac}       
\usepackage{graphicx}
\usepackage{hyperref}
\usepackage{microtype} 
\usepackage{tikz}
\usepackage{amsmath}
\usepackage{amssymb}
\usepackage{algorithm}
\usepackage[noend]{algpseudocode}

\usepackage{xcolor}         

\usepackage{geometry}

\newtheorem{theorem}{Theorem}
\newtheorem{prob}{Problem}

\newtheorem{assume}{Assumption}

\newtheorem{remark}{Remark}%

\newtheorem{definition}{Definition}

\newcommand{\Real}{\mathbb{R}}
\newcommand{\Natural}{\mathbb{N}}

\newcommand{\bfv}{\boldsymbol{v}}
\newcommand{\bfc}{\boldsymbol{c}}
\newcommand{\bfx}{\boldsymbol{x}}
\newcommand{\bfu}{\boldsymbol{u}}

\newcommand{\bfh}{\boldsymbol{h}}
\newcommand{\bfz}{\boldsymbol{z}}
\newcommand{\bfw}{\boldsymbol{w}}

\newcommand{\bfy}{\boldsymbol{y}}

\definecolor{mycolor}{rgb}{0.7,0.3,0.3}
\newcommand{\red}{\textcolor{mycolor}}

\title{The Feasibility and Inevitability of Stealth Attacks}
	
	\author{ 
  Ivan Y.~Tyukin\thanks{Corresponding Author} \\
  King's College London\\
  Strand, London, WC2R 2LS, UK \\
  \texttt{ivan.tyukin@kcl.ac.uk} \\
     \and
   Desmond J. Higham \\
  University of  Edinburgh\\
   Edinburgh, EH9 3FD, UK\\
   \texttt{d.j.higham@ed.ac.uk} \\
      \and
   Alexander Bastounis \\
  University of  Edinburgh\\
   Edinburgh, EH9 3FD, UK\\
   \texttt{abastoun@exseed.ed.ac.uk} \\
      \and
   Eliyas Woldegeorgis \\
  University of Leicester\\
  Leicester, LE1 7RH, UK \\
  \texttt{ew212@le.ac.uk} \\
  \and
   Alexander N. Gorban \\
  University of Leicester\\
  Leicester, LE1 7RH, UK \\
  \texttt{a.n.gorban@le.ac.uk} \\
}

\begin{document}

\date{}
\maketitle	
	
\begin{abstract} We develop and study new adversarial perturbations that enable an attacker to gain control over decisions in generic Artificial Intelligence (AI) systems including deep learning neural networks. In contrast to adversarial data modification, the attack mechanism we consider here involves  alterations to the AI system itself. Such a \emph{stealth attack} could be conducted by a mischievous, corrupt or disgruntled member of a  software development team. It could also be made by those wishing to exploit a ``democratization of AI'' agenda, where network architectures and trained parameter sets are shared publicly.  We develop a range of new implementable attack strategies with accompanying analysis, showing that with high probability  a stealth attack can be made transparent, in the sense that system performance is unchanged on a fixed validation set which is unknown to the attacker, while  evoking any desired output on a trigger input of interest.  The attacker only needs to have estimates of the size of the validation set and the spread of the AI's relevant latent space. In the case of deep learning neural networks, we show that a \emph{one neuron attack} is possible---a modification to the weights and bias associated with a single neuron---revealing a vulnerability arising from over-parameterization. We illustrate these concepts 
		using state of the art architectures on two standard
		image 
		data sets. Guided by the theory and computational results, we also propose strategies to guard against stealth attacks.
		
{\it Keywords}: Adversarial attacks, AI instability, concentration of measure, backdoor attacks, AI robustness
\end{abstract}

	
	\maketitle

	\section{Introduction}
	It is widely recognized that Artificial Intelligence (AI) systems can be vulnerable to 
	adversarial attacks \cite{szegedy2013intriguing}; that is, small, often imperceptible, perturbations that exploit instabilities.
	The danger of an attacker gaining control of an automated decision-making process is of particular concern in 
	high-stakes or safety-critical settings, including  medical imaging \cite{Fin_2020,GOB21},
	transport \cite{Wu_2020} and  textual analysis \cite{ebrahimi-etal-2018-hotflip}.
	The last few years have therefore seen an escalation in the design of both attack and defence strategies
	\cite{Attack_survey_2018,survey_sct_2020,Adv_train:2018}, and  recent work has considered the bigger question of whether 
	the existence of instabilities and successful attacks is inevitable \cite{Hansen20,shafahi2018adv, bastounis2021mathematics,tyukin2020adversarial,colbrook2022difficulty}.

	Work in this field has focused on adversarial perturbations to the training or input data, and may be classified as black-box,
	grey-box or white-box, depending on the level of information and access available to the attacker.
	We look at a different setting, which was introduced in \cite{tyukin2020adversarial} and is motivated further
	in subsection~\ref{sec:general_statement}.
	
	Here, the attacker is assumed to have control of the AI system, in the sense of being able to  edit code or change parameters. 
	We therefore consider attacks that  make perturbations to the system itself in an attempt to change the output on a specific input point of interest to the attacker {\it without any changes} in the AI's performance on a large, unknown, validation set and {\it without retraining}. We focus on two issues: (1) strategies for constructing such attacks and (2) theoretical analysis of the likelihood of success.

	In Section~\ref{sec:stealth} we present a general framework of open-box stealth attack and discuss additional related work. In Section \ref{sec:formal_statement_assumptions} we reveal {\it a new class of attacks} (Algorithms \ref{alg:plain} -- \ref{alg:constrained}) and establish probability bounds of their success (Theorems \ref{thm:plain} -- \ref{thm:concentration_collapse}). 
	
	Algorithm \ref{alg:plain}, and the corresponding Theorem \ref{thm:plain}, consider the case when the Attacker's choice of triggers is limited only by the requirement that the triggers' feature representations stay within a ball containing the Owner's validation set (choosing triggers with feature representations outside of this ball always results in successful attacks). In this case, we show  that the likelihood of  attack success can be made {\it arbitrarily} close to one, {\it for any fixed dimension} of the model's latent space (Theorem \ref{thm:plain}, Remark \ref{rem:probability_of_success}). This is a significant departure from the previous state of knowledge, as success likelihoods for such attacks were thought to be limited by dimension \cite{tyukin2020adversarial}. 	To establish these high probabilities of success, the attack must be executed with arbitrarily high accuracy and the model must satisfy appropriate reachability conditions \cite{survey_sct_2020}.
	
	Algorithm \ref{alg:incremental} and Theorem~\ref{thm:incremental} relate to approaches that enable attack triggers to be camouflaged as legitimate data by requesting that the Attacker's triggers produce latent representations within some given neighborhood of those corresponding to specified inputs.
	
	Algorithm \ref{alg:constrained} and Theorem \ref{thm:concentration_collapse} consider the case where the Attackers' capability to change or explore feature spaces of the model is constrained to some finite number of attributes or a smaller-dimensional subspace. The case is motivated by the ideas from \cite{BH22} where the authors proposed methods to generate adversarial perturbations confined to smaller-dimensional subspaces of the original input space (in contrast to the stealth attack setting considered here). This scenario enables attacks for models with sparse data representations in latent spaces. Remarkably, these constrained attacks may have significant probability  of success even when the accuracy of their implementation is relatively low; see Theorem~\ref{thm:concentration_collapse}.
	
	Section~\ref{sec:experiments} presents experiments which illustrate the application of the theory to realistic settings and demonstrates the strikingly likely feasibility of {\it one neuron attacks}---which alter weights of just a single neuron.  Section~\ref{sec:conclusion} concludes with recommendations on how vulnerabilities we exposed in this work can be mitigated by model design practices. 
	Proofs of the theorems can be found in the Appendix, along with extra algorithmic details and computational results.

	\section{Stealth attacks}\label{sec:stealth}

	\subsection{General framework}\label{sec:general_statement}
	
	Consider a generic AI system, a map
	\begin{equation}\label{eq:classification_map}
		\mathcal{F}:{\mathcal{U}}\rightarrow \Real
	\end{equation}
	producing some decisions on its outputs in response to an input from $\mathcal{U}\subset\Real^m$. The map $\mathcal{F}$ can define input-output relationships for an entire deep neural network or some part (a sub-graph),  an ensemble of networks, a tree, or a forest.  For the purposes of our work, the AI system's specific task is not relevant and can include  classification, regression, or density estimation. 
	
	In the classification case, if there are multiple output classes then 
	we regard (\ref{eq:classification_map}) as representing the output component of interest---we consider changes to $\mathcal{F}$ that do not affect any other output components; this is the setting in which our computational experiments are conducted. The flexibility for stealth attacks to work independently of the choice of output component is a key feature of our work.
	
\begin{figure}
		\centering
		A -- Classical adversarial attacks
		\vspace{2mm}
		
		\includegraphics[width=0.95\textwidth]{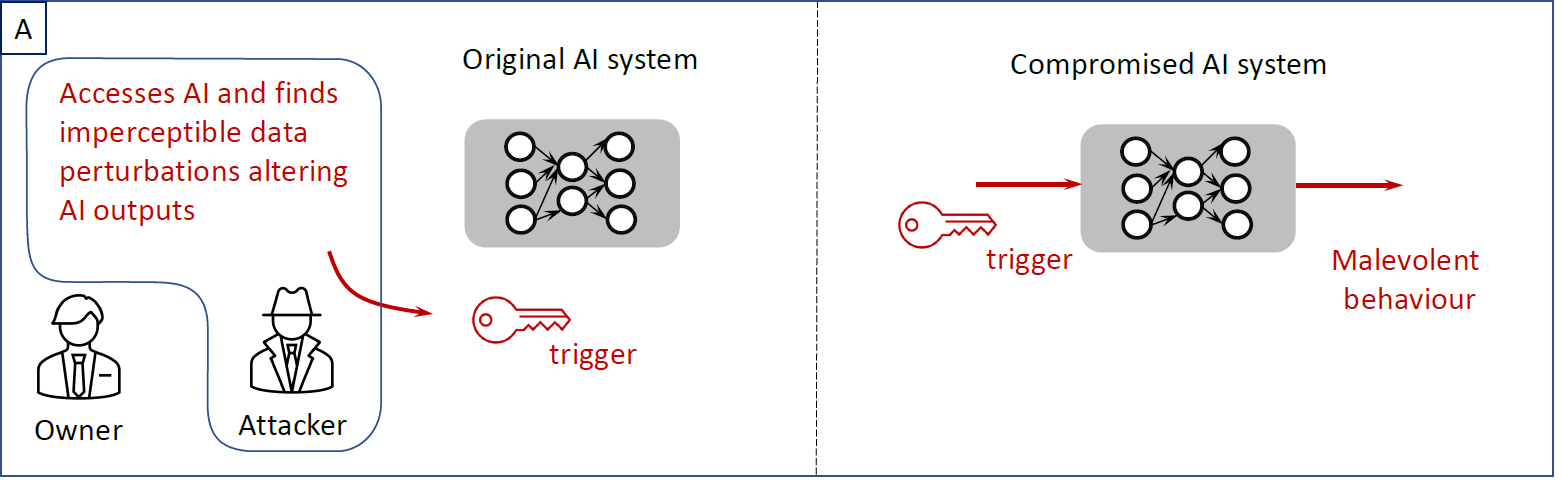}
		
		B -- Data poisoning attacks
		\vspace{2mm}
		
		\includegraphics[width=0.95\textwidth]{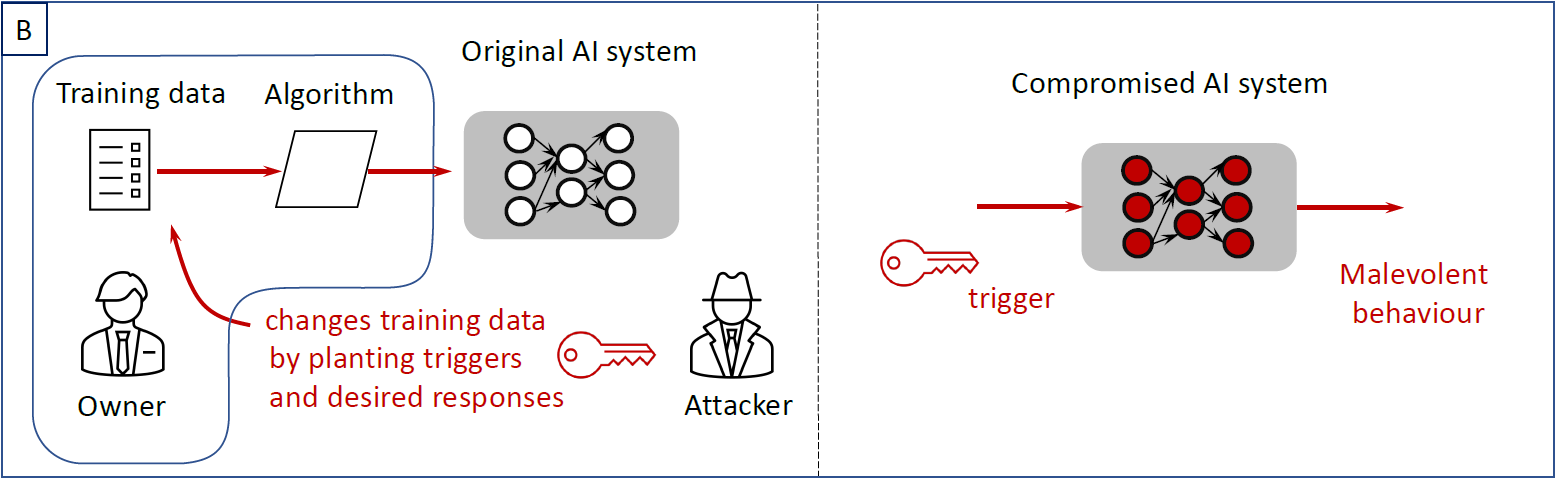}
		
		C -- Stealth attacks
		\vspace{2mm}
		
		\includegraphics[width=0.95\textwidth]{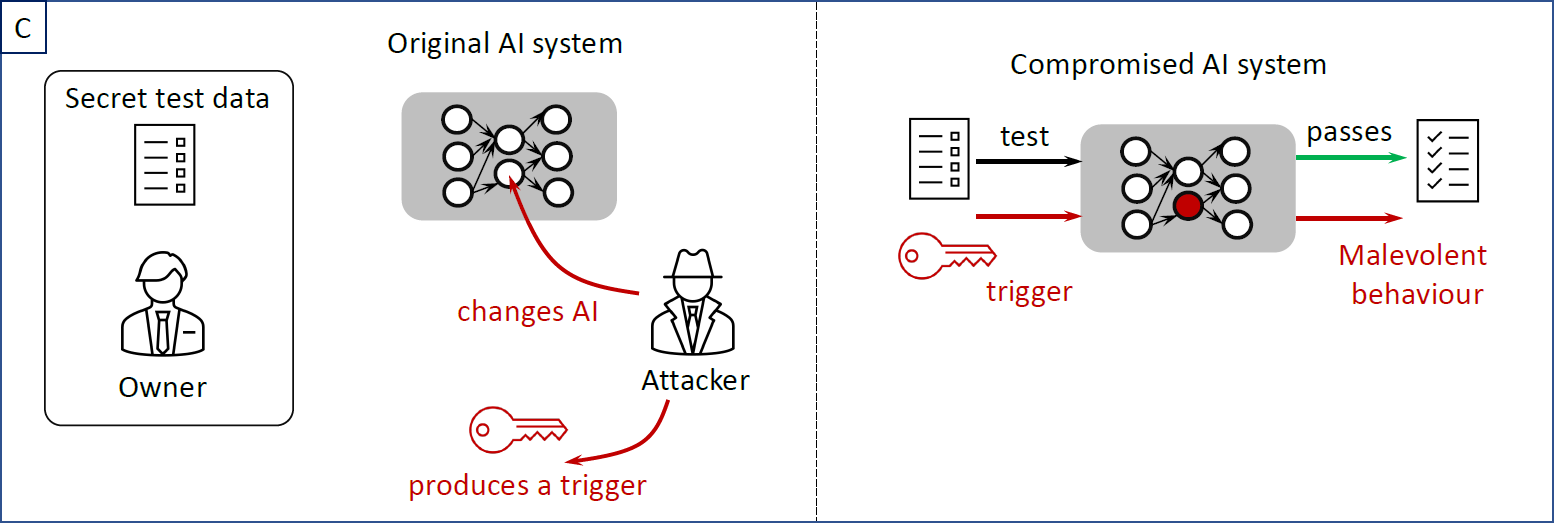}
		\caption{General schemes of adversarial (panel A), data poisoning (panel B), and stealth attacks (panel C).}\label{fig:stealth_attack}
	\end{figure}

	The AI system has an Owner operating the AI. An Attacker wants to take advantage of the AI by forcing it to make decisions in their favour. Conscious about security, the Owner created a \emph{validation set} which is kept secret. The validation set is a list of input-output pairs produced by the uncompromised  system (\ref{eq:classification_map}). The Owner can monitor security by
	checking that the AI reproduces these outputs. 
	Now, suppose that the Attacker has access to the AI system but not  the validation set. The phrase \emph{stealth attack}
	was used in \cite{tyukin2020adversarial}
	to describe the circumstance where the Attacker
	chooses a \emph{trigger input} and 
	\emph{modifies the AI} so that: 
	\begin{itemize}
		\item the Owner could not detect this modification by testing on the validation set,
		\item on the the trigger input the modified AI produces the output desired by the Attacker. 
	\end{itemize}
	Figure~\ref{fig:stealth_attack}, panel C, gives a schematic 
	representation of this setup. The setup is different from other known attack types such as adversarial attacks in which the Attacker exploits access to AI to compute imperceptible input perturbations altering AI outputs 
	(shown in Figure~\ref{fig:stealth_attack}, panel A), and data poisoning attacks (Figure~\ref{fig:stealth_attack}, panel B) in which the Attacker exploits access to AI training sets to plant triggers directly.
	
	We note that the stealth attack setting is relevant 
	to the case of a 
	corrupt, disgruntled or mischievous 
	individual who is 
	a member of a 
	software development team that is creating an AI system, 
	or who has an IT-related role where the system is deployed. 
	In this scenario, the attacker will have access to the AI system
	and, for example, in the case of a deep neural network, may choose to alter the weights, biases or architecture.
	The scenario is also pertinent in contexts where
	AI systems are exchanged between parties, such as 
	\begin{itemize}
		\item 
		``democratization of AI'' \cite{democ_ai},
		where copies
		of large-scale models and parameter sets are made available across multiple 
		public domain repositories,
		\item transfer learning \cite{Transfer20}, where an existing, previously trained tool is used as a starting point in a new application domain,
		\item outsourced cloud computing, where a third party service conducts  
		training \cite{Outsource17}.
	\end{itemize}
	
	\subsection{Formal definition of stealth attacks}
	
	Without loss of generality, it is convenient to represent the initial general map (\ref{eq:classification_map}) as a composition of two maps, $F$ and $\Phi$:
	\begin{equation}\label{eq:classification_map_composition}
		\mathcal{F}=F \circ \Phi,
		\quad \text{where}
		\quad 
		F: \ \Real^n \rightarrow \Real, \ \Phi:\mathcal{U}\rightarrow\Real^n.
	\end{equation}
	The map $\Phi$ defines general {\it latent} representation of inputs from $\mathcal{U}$, whereas the map $F$ can be viewed as a {\it decision-making} part of the AI system. In the context of deep learning models, latent representations can be outputs of hidden layers in deep learning neural networks, and decision-making parts could constitute operations performed by fully-connected and softmax layers at the end of the networks.  If $\Phi$ is an identity map then setting $F=\mathcal{F}$ brings us to the initial  case (\ref{eq:classification_map}). 
	
	An additional advantage of the compositional representation (\ref{eq:classification_map_composition}) is that it enables explicit modelling of the {\it focus of the adversarial attack}---a part of the AI system subjected to adversarial modification. This part will be modelled by be the map $F$. 
	
	A perturbed, or attacked,  map $F_a$ is defined as
	\begin{equation}\label{eq:adversarial_map}
		\begin{split}
			& {F}_{a}:{\Real^n}\times \Theta \rightarrow \Real, \\
			& {F}_{a}(\cdot,\boldsymbol{\theta})={F}(\cdot) + \mathfrak{A}(\cdot,\boldsymbol{\theta}),
		\end{split}
	\end{equation}
	where the term
	$
	\mathfrak{A}: {\Real^n}\times \Theta \rightarrow \Real
	$
	models the {\it effect} of an adversarial perturbation, and $\Theta\subset \Real^m$ is a set of relevant parameters.
	Such adversarial perturbations  could take different forms including modification of parameters of the map $F$ and physical addition or removal of components involved in computational processes in the targeted AI system. In neural networks the parameters are the weights and biases of a neuron or a group of neurons, and the components are neurons themselves. As we shall see later, a particularly instrumental case occurs when the term $\mathfrak{A}$ is {just a single} Rectified Linear Unit (ReLU function),   \cite{hahnloser2000digital},
	and $\boldsymbol{\theta} = (\bfw,b)$ represents the weights and bias, 
	\begin{equation}\label{eq:ReLU}
		\mathfrak{A}(\cdot,  \bfw,b )  = D \, \mathrm{ReLU}(\langle \cdot,\bfw \rangle -b)
		\quad 
		\text{where}
		\quad 
		\mathrm{ReLU}(s)=\max \{s,0\},
	\end{equation}
	or a sigmoid (see \cite{gorban1996neural}, \cite{HHdl2019} for further information on activation functions of different types)
	\begin{equation}\label{eq:sigmoid}
		\mathfrak{A}(\cdot,\bfw,b)= D \, \sigma(\langle \cdot,\bfw\rangle-b),
		\quad \text{where} \quad \sigma(s)=\frac{1}{1+\exp(- s)},
	\end{equation}
	where, $D\in\Real$ is a constant gain factor.

	Having introduced the relevant notation, we are now ready to provide a formal statement of the problem of stealth attacks introduced in Section \ref{sec:general_statement}.

	\begin{prob}[$\varepsilon$-$\Delta$ Stealth Attack on $\mathcal{F}$]\label{problem:stealth_plain}  \normalfont Consider a classification map $\mathcal{F}$ defined by (\ref{eq:classification_map}), (\ref{eq:classification_map_composition}). Suppose that an owner of the AI system has a finite  validation (or verification) set $\mathcal{V}\subset{\mathcal{U}}$. 
		The validation set $\mathcal{V}$ is kept secret and is assumed to be {\it unknown} to an attacker. The cardinality of $\mathcal{V}$ is bounded from above by some constant $M$, and this bound is known to the attacker.
		
		Given $\varepsilon\geq 0$ and $\Delta>0$, a successful 
		$\varepsilon$-$\Delta$ stealth attack takes place
		if the attacker modifies the map $F$ in $\mathcal{F}$ and replaces it by ${F}_a$ constructed such that for some $\bfu'\in \mathcal{U}$, known to the attacker but unknown to the owner of the map $\mathcal{F}$, the following properties hold:
		\begin{equation}\label{eq:adversarial_constraints}
			\begin{split}
				&|F\circ\Phi(\bfu)- F_a\circ \Phi(\bfu) |\leq \varepsilon \ \mbox{for all} \ \bfu\in\mathcal{V}\\
				&| {F}\circ\Phi(\bfu') - {F}_a\circ \Phi(\bfu') | \geq \Delta.
			\end{split}
		\end{equation}
	\end{prob}
	In words, when $F$ is perturbed to ${F}_a$ the output is changed by no more than $\varepsilon$ on the validation set, but is changed
	by at least $\Delta$ on the trigger, $\bfu'$.
	We note that this definition does not require any notion of whether the classification of $u$ or $u'$ is ``correct.'' In practice we are interested in cases where 
	$\Delta$ is sufficiently large for 
	$u'$ to be assigned to different classes
	by the original and perturbed maps.
	
	
	\begin{remark}[The target class for the trigger $\bfu'$]\label{rem:multiclass} \normalfont
		Note that the above setting can be adjusted to fit a broad range of problems, including general multi-class problems. Crucially, the stealth attacks proposed in this paper allow the attacker to choose which class the modified AI system $F_a$ predicts for the trigger image $\bfu'$. We illustrate these capabilities with examples in Section \ref{sec:experiments}.
	\end{remark}
	
	\subsection{Related work}\label{sec:related_work}
	
	{\it Adversarial attacks}. A broad range of methods aimed at revealing vulnerabilities of state-of-the art AI, and deep learning models in particular, to adversarial inputs has been proposed to date (see e.g. recent reviews  \cite{survey_sct_2020,REN2020346}). The focus of this body of work has been primarily on perturbations to signals/data processed by the AI. In contrast to this established framework, here we explore possibilities to determine and implement small {\it changes to AI structure} and {\it without retraining}.
	
	{\it Data poisoning}. Gu et al \cite{gu2017badnets} (see also \cite{BIGGIO2018317} and references therein, and \cite{manoj2021excess} for explicit upper bounds on the volume of poisoned data that is sufficient to execute these attacks) showed how malicious data poisoning occurring, for example, via outsourcing of training to a third party, can lead to {\it backdoor} vulnerabilities. 
	Performance of the modified model on the validation set, however, is not required to be perfect. A data poisoning attack is deemed successful if the model's performance on the validation set is within some margin of the user's expectations.     
	
	The data poisoning scenario is different from our setting in two fundamental ways. First, in our case the attacker can maintain performance of the perturbed system on an unknown validation set within {\it arbitrary small} or, in case of ReLU neurons, {\it zero} margins. Such {\it imperceptible} changes on unknown validation sets is a signature characteristic of the threat we have revealed and studied.  Second, the attacks we analysed {\it do not require any retraining}. 
	

	{\it Other stealth attacks.} Liu et al \cite{stealth_hardware:2018} proposed a mechanism, $SIN^2$, whereby a service provider gains control over their customers' deep neural networks through the provider's specially designed APIs. In this approach, the provider needs to first  plant malicious information in higher-precision bits (e.g. bits $16$ and higher) of the network's weights. When an input trigger arrives the provider extracts this information via its service's API and performs malicious actions as per instructions encoded. 
	
	In contrast to this approach, stealth attacks we discovered {\it do not require} any special computational environments. After an attack is planted, it can be executed in fully secure and trusted infrastructure. In addition, our work reveals further concerns about how easily a malicious service provider can implement stealth attacks in hostile environments \cite{stealth_hardware:2018} by swapping bits in the mantissa of weights and biases of a single neuron (see Figure~\ref{fig:stealth_planting} for the patterns of change) at will.

	Our current work is a significant advancement from the preliminary results presented in \cite{tyukin2020adversarial}, both in terms of algorithmic detail and theoretical understanding of the phenomenon. First, the vulnerabilities we reveal here are much more severe. Bounds on the probability of success for attacks in \cite{tyukin2020adversarial} are constrained by $1-M 2^{-n}$. Our results show that under the same assumptions (that input reachability \cite{survey_sct_2020} holds true), the probabilities of success for attacks generated by Algorithms \ref{alg:plain}, \ref{alg:incremental} can be made arbitrarily close to one (Remark \ref{rem:probability_of_success}). Second, we explicitly account for cases when input reachability holds only up to some accuracy, through parameters $\alpha,\delta$ (Remark~\ref{rem:parameters}). Third, we present concrete algorithms, scenarios and examples of successful exploitation of these new vulnerabilities,
	including the case of one neuron attacks
	(Section \ref{sec:experiments}, Appendix; code is available in \cite{ExampleCode}).

	\section{New stealth attack algorithms}\label{sec:formal_statement_assumptions}
	
	In this section we introduce two new algorithms for generating stealth attacks on a generic AI system. These algorithms return a ReLU or a sigmoid neuron realizing an adversarial perturbation $\mathfrak{A}$.  Implementation of these algorithms will rely upon some mild additional information about the unknown validation set $\mathcal{V}$.  In particular, we request that latent representations $\Phi(\bfu)$ for all $\bfu\in\mathcal{V}$ are located within some ball $\mathbb{B}_n(0,R)$ whose radius $R$ is {\it known} to the attacker. We state this requirement in Assumption \ref{assume:data_ball} below.
	
	\begin{assume}[Latent representations of the validation data $\mathcal{V}$]\label{assume:data_ball} There is an $R>0$, known to the attacker, such that
		\begin{equation}\label{eq:data_ball}
			\Phi(\bfu) \in \mathbb{B}_n(0,R) \ \mbox{for all} \ \bfu\in\mathcal{V}.
		\end{equation}
	\end{assume}
	
	Given that the set $\mathcal{V}$ is finite, infinitely many such balls exist. Here we request that the attacker knows just a single value of $R$ for which  (\ref{eq:data_ball}) holds. The value of $R$ does not have to be the smallest possible.
	
	We also suppose that the feature map $\Phi$ in (\ref{eq:classification_map_composition}) satisfies an appropriate reachability condition (cf. \cite{survey_sct_2020}), based on the following definition.  
	
	\begin{definition}[$\upsilon$-input reachability of the classifier's latent space] Consider a function $f:\mathcal{U}\rightarrow \Real^n$. A set $\mathcal{S}\subset\Real^n$ is $\upsilon$-input reachable for the function $f$ if for any $\bfx\in\mathcal{S}$ there is an $\bfu(\bfx)\in\mathcal{U}$ such that $\|f(\bfu) - \bfx\|\leq \upsilon$.
	\end{definition}
	
	In what follows we will assume existence of some sets, namely $n-1$ spheres, in the classifier's latent spaces which are input reachable for the map $\Phi$. Precise definitions of these sets will be provided in  our formal statements.
	
	\begin{remark}\label{rem:input_reachability}\normalfont The requirement that the classifier's latent space contains specific sets which are $\upsilon$-input reachable for the map $\Phi$ may appear restrictive. If, for example, feature maps are vector compositions of ReLU and affine mappings (\ref{eq:ReLU}), then  several components of these vectors might be equal to zero in some domains.  At the same time, one can always search for a new feature map
		\begin{equation}\label{eq:modified_feature_maps}
			\tilde\Phi: \ \tilde{\Phi}(\bfu) = T \Phi(\bfu), \ T\in\Real^{d \times n}, 
		\end{equation}
		where the matrix $T\in\Real^{d\times n}$, $d\leq n$ is chosen so that relevant domains in the latent space formed by $\tilde\Phi$  become  $\upsilon$-input reachable.  As we shall see later in Theorems~\ref{thm:plain} and \ref{thm:incremental}, these relevant domains are  spheres $\mathbb{S}_{d-1}(\bfc,\delta R)$, $\delta\in(0,1]$, where $\bfc\in\Real_d$ is some given reference point. 
		
		If the feature map $\Phi:\mathcal{U}\rightarrow \Real^n$ with $\mathcal{U}\subset \Real^m$ is differentiable and non-constant, and the set $\mathcal{U}$ has a non-empty interior, $\mathrm{Int}(\mathcal{U})$, then matrices $T$ producing $\upsilon$-input reachable feature maps $\tilde{\Phi}$ in some neighborhood of a point from $\mathrm{Int}(\mathcal{U})$ can be determined as follows. Pick a target input $\bfu_0\in\mathrm{Int}(\mathcal{U})$, then
		\[
		\Phi(\bfu)=\Phi(\bfu_0)+ J (\bfu-\bfu_0) + o(\|\bfu-\bfu_0\|),
		\]
		where $J$ is the Jacobian of $\Phi$ at $\bfu_0$. Suppose that $\mathrm{rank}(J)=d$, $d>0$, and let
		\[
		J= U \left(\begin{array}{cc} \Sigma & 0_{d\times (m-d)} \\ 0_{(n-d)\times d} & 0_{(n-d)\times (m-d)} \end{array}\right) V
		\]
		be the singular value decomposition of $J$, where $\Sigma$ is a $d\times d$ diagonal matrix containing non-zero singular values of $J$ on its main diagonal, and $U\in \Real^{n\times n}$, $V\in\Real^{m\times m}$ are unitary matrices. If $m=d$ or $n=d$ then the corresponding zero matrices in the above decomposition are assumed to be empty. Setting 
		\[
		T = \left( I_d \ 0_{d\times (n-d)} \right) U^{\mathsf{T}}
		\]
		ensures that for any arbitrarily small $\upsilon>0$ there is an $r(\upsilon,\bfu_0)>0$ such that the sets $\mathbb{S}_{d-1}(\tilde{\Phi}(\bfu_0), \rho)$, $\rho\in[0,r(\upsilon,\bfu_0))$ are $\upsilon$-input reachable for the function $\tilde{\Phi}=T \Phi$.   
		
		Note that the same argument applies to maps $\Phi$ which are differentiable only in some neighborhood of $\bfu_0\in\mathrm{Int}(\mathcal{U})$. This enables the application of this approach for producing $\upsilon$-input reachable feature functions to maps involving compositions of ReLU functions. In what follows we will use the symbol $n$ to denote the dimension of the space where $\Phi$ maps the input into assuming that the input reachability condition holds for this feature map.
		
		Linearity of $\tilde{\Phi}$ in $T$ enables to preserve the structure of perturbations (\ref{eq:ReLU}), (\ref{eq:sigmoid}) so that they remain, in effect, just a single additional neuron. We used the latter property in our computational experiments (see Remark \ref{rem:perturbation_dimension} in Section \ref{sec:choosing_triggers}) to generate examples of stealth attacks.

	\end{remark}
	
	\subsection{Target-agnostic stealth attacks}
	
	Our first stealth attack algorithm is presented in Algorithm \ref{alg:plain}. The algorithm produces a modification of the relevant part $F$ of the original AI system that is implementable by a single ReLU or sigmoid function.  Regarding the {\it trigger input}, $\bfu'$, the algorithm relies on another process (see step 3).
	
	In what follows, we will denote this process as an auxiliary algorithm $\mathcal{A}_{0}$ which, for the map $\Phi$, given $R$, $\delta\in(0,1]$,$\gamma\in(0,1)$, $\upsilon<\delta$, $\delta+\upsilon \leq \gamma^{-1}$, and any $\bfx\in\mathbb{S}_{n-1}(0,\delta)$, returns a solution of the following constrained search problem:
	\begin{equation}\label{eq:finding_trigger_1}
		\begin{split}
			& \mathcal{A}_0(\Phi,\bfx,R,\gamma,\upsilon): \quad \text{find} \quad \bfu\in\mathcal{U}   
			\quad 
			\text{such that}  \quad 
			\left\| \Phi(\bfu) R^{-1} \right\| \leq \gamma^{-1}, \  \left\|\Phi(\bfu) R^{-1} - \bfx \right\| <  \upsilon. 
		\end{split}
	\end{equation}
	Observe that $\upsilon R$-input reachability, $\upsilon<\delta$, of the set $\mathbb{S}_{n-1}(0,R\delta)$ for the map $\Phi$ together with the choice of $\delta,\upsilon$, and $\gamma$ satisfying $\delta+\upsilon\leq \gamma^{-1}$  ensure existence of a solution to the above problem for every $\bfx\in \mathbb{S}_{n-1}(0,\delta)$. 
	
	Thorough analysis of computability of solutions of (\ref{eq:finding_trigger_1}) is outside of the scope of the work (see \cite{bastounis2021extended} for an idea of the issues involved computationally in solving optimisation problems). Therefore,  we shall assume that the auxiliary process $\mathcal{A}_0$ always returns a solution of (\ref{eq:finding_trigger_1}) for a choice of $\upsilon, \delta, \gamma$, and $R$. In our numerical experiments (see Section \ref{sec:experiments}), finding a solution of (\ref{eq:finding_trigger_1}) did not pose significant issues.
	

	With the auxiliary algorithm $\mathcal{A}_0$ available, we can now present Algorithm \ref{alg:plain}:

	\begin{algorithm}
		\caption{Single-neuron plain stealth attack}\label{alg:plain}
		\underline{Input}: $\delta\in(0,1]$,  $\gamma\in(0,1)$, $\alpha\in[0,1)$, $(1+\alpha)\delta \leq 1$, $\Delta,\varepsilon \geq 0$, a sigmoid or a ReLU function $g$, and $R$ satisfying (\ref{eq:data_ball}).
		
		\begin{algorithmic}[1]
			\Procedure{Adversarial stealth perturbation}{$\delta,\Delta,\varepsilon,g$}      
			\State  Draw a random vector $\bfx$ from the equidistribution in the sphere ${\mathbb{S}_{n-1}}(0,\delta)$, $\delta\in(0,1]$.
			\State Use algorithm $\mathcal{A}_0$ (see (\ref{eq:finding_trigger_1})) to generate an input $\bfu'\in\mathcal{U}$ such that $\bfx'=\Phi(\bfu')/{R}$ is within a $\alpha \delta$-distance from $\bfx$:
			\begin{equation}\label{eq:trigger_condition_1}
				\|\bfx'-\bfx\|\leq \alpha \delta.
			\end{equation}
			
			\State Set 
			\begin{equation}\label{eq:adversarial_solution}
				\mathfrak{A}\left(\cdot, \kappa \bfx' R^{-1}, b \right)=D g\left(\langle \cdot,\kappa \bfx'R^{-1} \rangle  - b\right), \quad
				\text{with} \quad 
				b=0.5\kappa \left(1+\gamma\right) \|\bfx'\|^2,
			\end{equation}
			where   $\kappa$ and $D$ are chosen so that
			\begin{equation}\label{eq:adversarial_solution_parameters}
				\begin{split}
					&D g \left(-0.5\kappa (1-\gamma) \|\bfx'\|^2 \right)  \leq \varepsilon \ \mbox{and} \ D g \left(0.5\kappa (1-\gamma) \|\bfx'\|^2 \right)  \geq \Delta.
				\end{split}
			\end{equation}
			
			$^\ast$Note that a choice of $\kappa, D$  so that (\ref{eq:adversarial_solution_parameters}) is satisfied is always possible.
			
			\EndProcedure
		\end{algorithmic}
		\underline{Output}: trigger $\bfu'$, weight vector $\bfw=\kappa\bfx'R^{-1}$,  bias $b=0.5\kappa \left(1+\gamma\right) \|\bfx'\|^2$, and output gain $D$ of the sigmoid or  ReLU function $g$.
	\end{algorithm}
	
	The performance of Algorithm \ref{alg:plain} in terms of the probability of producing a successful stealth attack is characterised by Theorem \ref{thm:plain} below.

	\begin{theorem}\label{thm:plain}  Let Assumption~\ref{assume:data_ball} hold. Consider Algorithm \ref{alg:plain}, and let  $\gamma\in(0,1)$, $\alpha \in [0,1)$, and $\delta\in (0,1]$ be such that $(1+\alpha)\delta\leq 1$ and $\gamma(1-\alpha)\delta > \alpha$. Additionally, assume that the set $\mathbb{S}_{n-1}(0,R\delta)$ is $\delta \alpha R$-input reachable for the map $\Phi$. Then
		\begin{equation}\label{eq:algorithm_parameters}
			\varphi(\gamma,\delta,\alpha):=\cos(\arccos(\gamma (1-\alpha)\delta) + \arccos((1-\alpha^2)^{1/2}))>0
		\end{equation}
		and the probability $P_{a,1}$ that Algorithm \ref{alg:plain} returns a successful $\varepsilon$-$\Delta$ stealth attack is greater than or equal to
		\begin{equation}\label{eq:probability_plain}
			1-M  \pi^{-1/2} \frac{\Gamma\left(\frac{n}{2}\right)}{\Gamma\left(\frac{n-1}{2}\right)} \int_{0}^{\arccos{( \varphi(\gamma,\delta,\alpha) )}} \sin^{n-2}(\theta)d\theta.
		\end{equation}
		In particular, $P_{a,1}$ is greater than or equal to
		\begin{equation}\label{eq:probability_plain_bound}
			1 - M  \frac{1}{2 \pi^{\frac{1}{2}}} \frac{\Gamma\left(\frac{n}{2}\right)}{\Gamma\left(\frac{n}{2}+\frac{1}{2}\right)} \frac{1}{\varphi(\gamma,\delta,\alpha)} \left(1-\varphi(\gamma,\delta,\alpha)^2\right)^{\frac{n-1}{2}}.
		\end{equation}
	\end{theorem}

	\begin{remark}[Choice of parameters]\label{rem:parameters} \normalfont The three parameters $\alpha,\delta$ and $\gamma$ should be chosen to strike a balance between attack success, compute time, and the size of the weights that the stealth attack creates. In particular, from the perspective of attack success \eqref{eq:probability_plain} is optimized when $\gamma$ and $\delta$ are close to $1$ and $\alpha$ is close to $0$. However, $\alpha$ and $\delta$ are restricted by the exact form of the map $\Phi$: they need to be chosen so that the set $\mathbb{S}_{n-1}(0,R\delta)$ is $\delta \alpha R$-input reachable for $\Phi$. Furthermore, the speed of executing $\mathcal{A}_0$ is also influenced by these choices with the complexity of solving \eqref{eq:trigger_condition_1} increasing as $\alpha$ decreases to $0$. The size of the weights chosen in \eqref{eq:adversarial_solution_parameters} is influenced by the choice of $\gamma$ so that the $L_2$ norm of the attack neuron weights grows like $O((1-\gamma)^{-1})$ for sigmoid $g$. Finally, the condition $(1+\alpha)\delta \leq 1$ ensures that the chosen trigger image $\bfu'$ has $\|\Phi(\bfu')\| \leq R$.
	\end{remark}
	
	\begin{remark}[Determinants of success and vulnerabilities]\label{rem:probability_of_success} \normalfont Theorem \ref{thm:plain} establishes explicit connections between intended parameters of the trigger $\bfu'$ (expressed by $\Phi(\bfu')/R$ which is to be maintained within $\delta\alpha$ from $\bfx$), dimension $n$ of the AI's {\it latent} space, accuracy of solving (\ref{eq:finding_trigger_1}) (expressed by the value of $\alpha$---the smaller the better),  design parameters $\gamma\in(0,1)$ and $\delta\in(0,1]$, and {\it vulnerability} of general AI systems to stealth attacks.

		In general, the larger the value of $n$ for which solutions of (\ref{eq:finding_trigger_1}) can be found without increasing the value of $\alpha$, the higher  the probability that the attack produced by Algorithm \ref{alg:plain} is successful. Similarly, for a given success probability bound, the larger the value of $n$, the smaller the value of $\gamma$ 
		required, allowing stealth attacks to be implemented with smaller weights $\bfw$ and bias $b$. At the same time, if no explicit restrictions on $\delta$ and the weights are imposed then  Theorem \ref{thm:plain} suggests that if one picks $\delta=1$, $\gamma$ sufficiently close to $1$, and $\alpha$ sufficiently close to $0$, subject to finding an appropriate solution of (\ref{eq:finding_trigger_1}) (c.f. the notion of reachability \cite{survey_sct_2020}), one can  
		create stealth attacks whose probabilities of success  {\it can be made arbitrarily close to $1$ for any fixed $n$}. Indeed, if $\alpha=0$ then the right-hand side of (\ref{eq:probability_plain_bound}) becomes
		\begin{equation}\label{eq:attack_probability_extreme}
			1 - M  \frac{1}{2 \pi^{\frac{1}{2}}} \frac{\Gamma\left(\frac{n}{2}\right)}{\Gamma\left(\frac{n}{2}+\frac{1}{2}\right)} \frac{1}{\gamma \delta} \left(1-(\gamma \delta)^2\right)^{\frac{n-1}{2}}
		\end{equation}
		which, for any fixed  $M$, $n$, can me made arbitrarily close to $1$ by an appropriate choice of $\gamma\in(0,1)$ and $\delta\in(0,1]$.
	\end{remark}
	
	
	\begin{remark}[Arbitrary output sign and target class]\label{rem:sign} 
	
	\normalfont Although Problem \ref{problem:stealth_plain} does not impose any requirements on the sign of  $\mathfrak{A}\left(\Phi(\bfu'),\bfw,b\right)$, 
		this quantity 
		can be made positive or negative through the choice of $D$. Note that the target class can be arbitrary too (see Remark \ref{rem:multiclass}).
	\end{remark}
	
	\begin{remark}[Precise and zero-tolerance attacks with ReLU units]\label{rem:ReLU}  \normalfont One of the original requirements for a stealth attack is to produce a response to a trigger input $\bfu'$ such that $| {F}\circ\Phi(\bfu') - {F}_a\circ \Phi(\bfu')|$ exceeds an a-priory given value $\Delta$. However, if the attack is implemented with a ReLU unit then one can select the values of $\bfw,b$ so that  $| {F}\circ\Phi(\bfu') - {F}_a\circ \Phi(\bfu')|$ exactly equals $\Delta$. Indeed for a ReLU neuron, condition (\ref{eq:adversarial_solution_parameters}) reduces to
		\[
		0.5 D  \left(\kappa (1-\gamma) \|\bfx'\|^2 \right)  \geq \Delta.
		\]
		Hence picking $\kappa=2 {\Delta}\left((1-\gamma) \|\bfx'\|^2\right)^{-1}$ and $D=1$ results in the desired output. 
		
		Moreover, ReLU units produce adversarial perturbations $\mathfrak{A}$ with $\varepsilon=0$. These {\it zero-tolerance} attacks, if successful, do not change the values of  $\mathcal{F}$ on the validation set $\mathcal{V}$ and as such are completely undetectable on $\mathcal{V}$. 
	\end{remark}
	
	\begin{remark}[Hiding adversarial attacks in redundant structures]\label{rem:hiding} \normalfont  Algorithm \ref{alg:plain} (and its input-specific version, Algorithm~\ref{alg:incremental}, below) implements adversarial perturbations by {\it adding} a single sigmoid or ReLU neuron. A question therefore arises, if one can ``plant'' or ``hide'' a stealth attack within an existing AI structure. Intuitively, over-parametrisation and redundancies in many deep learning architectures  should provide ample opportunities precisely for this sort of malevolent action. As we empirically justify in Section \ref{sec:experiments}, this is indeed possible.
		In these experiments, we looked at a neural network with $L$ layers. The  map $F$ corresponded to the last $k+1$ layers: $L-k, \dots, L$, $k\geq 1$. We looked for a neuron in layer $L-k$ whose output weights have the smallest $L_1$-norm of all neurons in that layer (see Appendix, Section \ref{sec:neuron_selection}). This neuron was then replaced with a neuron implementing our stealth attack, and its output weights were wired so that a given trigger input $\bfu'$ evoked the response we wanted from this trigger.
		This new type of one neuron attack may be viewed as the stealth version of a one pixel attack \cite{SVK17}.
		Surprisingly, this approach worked consistently well  across various randomly initiated instances of the same network. These experiments suggest a real non-hypothetical possibility of turning {\it a needle in a haystack} (a redundant neuron) into {\it a pebble in a shoe} (a malevolent perturbation).
	\end{remark}

	\begin{remark}[Other activation functions]\label{rem:activations} \normalfont 
		In this work, to be concrete we focus on the case of attacking with a ReLU or sigmoid 
		activation function. However, the algorithms and analysis readily extend to the case of
		any continuous activation function that is nonnegative and monotonically increasing.
		Appropriate pairs of matching leaky ReLUs also fit into this framework. 
	\end{remark}

	
	\subsection{Target-specific stealth attacks}
	
	The attack and the trigger $\bfu'\in\mathcal{U}$ constructed in Algorithm \ref{alg:plain} are ``arbitrary''
	in the sense that $x$ is drawn at  random. 
	As opposed to standard adversarial examples, they are not linked or targeting any specific input. Hence a question arises:  is it possible to create a {\it targeted} adversarial perturbation of the AI which is triggered by an input $\bfu'\in\mathcal{U}$  located in a vicinity of some specified input, $\bfu^\ast$?
	
	As we shall see shortly, this may indeed be possible through some modifications to Algorithm \ref{alg:plain}. For technical convenience and consistency, we introduce a slight reformulation of Assumption \ref{assume:data_ball}.
	
	\begin{assume}[Relative latent representations of the validation data $\mathcal{V}$]\label{assume:data_ball:targeted} Let $\bfu^\ast\in\mathcal{U}$ be a target input of interest to the attacker. There is an $R>0$, also known to the attacker, such that
		\begin{equation}\label{eq:data_ball_target}
			\Phi(\bfu)-\Phi(\bfu^\ast) \in \mathbb{B}_n(0,R) \ \mbox{for all} \ \bfu\in\mathcal{V}.
		\end{equation}
	\end{assume}
	We note that Assumption \ref{assume:data_ball:targeted} follows immediately from Assumption \ref{assume:data_ball} if a bound on the size of the latent
	representation $\Phi(\bfu^\ast)$ of the input $\bfu^\ast$ is known. Indeed, if $R'$ is a value of $R$ for which (\ref{eq:data_ball}) holds then (\ref{eq:data_ball_target}) holds whenever $R\geq R'+\|\Phi(\bfu^\ast)\|$.
	
	Algorithm \ref{alg:incremental} provides a recipe for creating such targeted attacks. 
	
	\begin{algorithm}
		\caption{Single-neuron targeted stealth attack}\label{alg:incremental} 
		Input: $\delta\in(0,1]$,  $\gamma\in(0,1)$, $\alpha\in[0,1)$, $\delta(\alpha+1) \leq 1$, $\Delta,\varepsilon \geq 0$, a sigmoid or a ReLU function $g$, a target input $\bfu^\ast\in\mathcal{U}$, and $R$ for which (\ref{eq:data_ball_target}) holds.
		\begin{algorithmic}[1]
			\Procedure{Targeted adversarial stealth perturbation}{$\delta,\alpha,D,\varepsilon,g$}      
			\State Draw a random vector $\bfx$ from the equidistribution in the sphere ${\mathbb{S}_{n-1}}(0,\delta)$, $\delta\in(0,1]$.
			\State Use the algorithm $\mathcal{A}_0$ to generate an input $\bfu'$ such that $\bfx'=(\Phi(\bfu')-\Phi(\bfu^\ast))/R$  satisfies (\ref{eq:trigger_condition_1}).
			\State Set 
			\begin{equation}
				\begin{split}
					& \mathfrak{A}\left(\cdot, \kappa {\bfx'}R^{-1}, b \right)=D g\left(\langle\cdot,\kappa \bfx'R^{-1}\rangle - b\right), \\
					& b=0.5\kappa \left(1+\gamma\right) \|\bfx'\|^2  + \kappa \langle\Phi(\bfu^\ast),\bfx'R^{-1}\rangle,
				\end{split}\nonumber
			\end{equation}
			where  $\kappa$ and $D$ are chosen as in (\ref{eq:adversarial_solution_parameters}).
			
			$^\ast$Note that such a choice is always possible.
			
			\EndProcedure
		\end{algorithmic}
		Output: trigger input $\bfu'$, weight vector $\bfw=\kappa\bfx'{R}^{-1}$,  bias $b=0.5\kappa ( 1+\gamma ) \|\bfx'\|^2 + \kappa \langle \Phi(\bfu^\ast),\bfx'R^{-1}\rangle$, and output gain $D$ of the sigmoid or a ReLU function $g$.
	\end{algorithm}
	
	Performance bounds for Algorithm~\ref{alg:incremental} can be derived in the same way as we have done in Theorem \ref{thm:plain} for Algorithm \ref{alg:plain}. Formally, we state these bounds in Theorem~\ref{thm:incremental} below.

	\begin{theorem}\label{thm:incremental} Assumption~\ref{assume:data_ball:targeted} hold. Consider Algorithm \ref{alg:incremental}, and let parameters $\gamma\in(0,1)$, $\alpha \in [0,1)$, and $\delta\in (0,1]$ be such that $\delta(1+\alpha) \leq 1$ and $\gamma(1-\alpha)\delta > \alpha$. Additionally, assume that the set $\mathbb{S}_{n-1}(\Phi(\bfu^\ast),R\delta)$ is $\delta \alpha R$-input reachable for the map $\Phi$.  
		
		Then the probability $P_{a,2}$ that Algorithm \ref{alg:incremental} returns a successful $\varepsilon$-$\Delta$ stealth attack is bounded from below by (\ref{eq:probability_plain}) and (\ref{eq:probability_plain_bound}).
	\end{theorem}

	Remarks~\ref{rem:parameters}--\ref{rem:activations} apply equally to Algorithm~\ref{alg:incremental}. In addition, the presence of a specified target input $\bfu^\ast$ offers extra flexibility and opportunities. An attacker may 
	for example have a list of potential target inputs. Applying Algorithm \ref{alg:incremental} to each of these inputs produces different triggers $\bfu'$ each with different possible values of $\alpha$. According to Remark \ref{rem:probability_of_success} (see also (\ref{eq:attack_probability_extreme})), small values of $\alpha$ imply higher probabilities that the corresponding attacks are successful. Therefore, having a list of target inputs and selecting a trigger with minimal $\alpha$ increases the attacker's chances of  success.


	\subsection{Attribute-constrained attacks and a ``concentrational collapse'' for  validation sets $\mathcal{V}$ chosen at random}\label{sec:other_bounds}

    The attacks developed and analysed so far can affect all attributes of the input data's latent representations. 
    It is also worthwhile to consider triggers whose deviation from targets in latent spaces is limited to only a few attributes. This
    may help to disguise triggers as legitimate data, an approach which was recently shown to be successful in the context of adversarial attacks on input data \cite{BH22}. Another motivation stems from more practical scenarios where the task is to find a trigger in the vicinity of the latent representation of another input in models with sparse coding. The challenge here is to produce attack triggers whilst retaining sparsity.  Algorithm \ref{alg:constrained} and the corresponding Theorem \ref{thm:concentration_collapse} are motivated by these latter scenarios.
	
	As before, consider an AI system with an input-output map (\ref{eq:classification_map_composition}). Assume that the attacker has access to a suitable real number $R>0$, an input $\bfu^\ast$, and a set of orthonormal vectors $h_1,h_2,\dotsc,h_{n_p} \in \Real^n$ which form the space in which the attacker can make perturbations. Furthermore, assume that both the validation set and the input $\bfu^\ast$ satisfy Assumption \ref{assume:concentration} below.
	
	\begin{assume}[Data model]\label{assume:concentration}  Elements $\bfu_1,\dots,\bfu_M$ are randomly chosen so that $\Phi(\bfu_1)$, $\dots$, $\Phi(\bfu_M)$ are random variables in $\Real^n$ and such that both of the following hold:
		
		\begin{enumerate}
			\item There is a $\bfc\in\Real^n$ such that $\bfu^*$, and (with probability $1$), $\bfu_1,\dots,\bfu_M$, are in the set:
			\begin{equation}\label{eq:support}
				\{\bfu\in\mathcal{U}\  | \ \|\Phi(\bfu)-\bfc\|\leq R/2 \}.
			\end{equation}

			\item There is a $C\geq 1$ such that for any $r\in(0,R/2]$ and  $\xi\in\mathbb{B}_n(c,R/2)$
			\begin{equation}\label{eq:exponent_concentration}
				\begin{split}
					&P(\Phi(\bfu_i)\in \mathbb{B}_n(\xi,r) )\leq C \left(\frac{ 2 r}{R}\right)^n. 
				\end{split}
			\end{equation}
		\end{enumerate}
	\end{assume} 
	

	Property (\ref{eq:support}) requires that the validation set is mapped into a ball centered at $\bfc$ and having a radius $R/2$ in the system's latent space, and (\ref{eq:exponent_concentration}) is a non-degeneracy condition restricting pathological concentrations. 
	
	Consider Algorithm \ref{alg:constrained}. As we show below, if the validation set satisfies Assumption \ref{assume:concentration} and the attacker uses Algorithm \ref{alg:constrained} then the probabilities of generating a successful attack may be remarkably high even if the attack triggers' latent representations deviate from those of the target in only few attributes.

	\begin{algorithm}
		\caption{Single-neuron targeted attribute-constrained stealth attack}\label{alg:constrained} 
		Input: $\delta\in(0,1]$,  $\gamma\in(0,1)$, $\alpha\in[0,1)$, $\delta(\alpha+1) \leq 1$, $\Delta,\varepsilon \geq 0$, a sigmoid or a ReLU function $g$, a target input $\bfu^\ast\in\mathcal{U}$, an $R$ for which Assumption \ref{assume:concentration} holds, an $n_p\in\Natural$, $2\leq n_p\leq n$, and a system of orthonormal vectors $\{\bfh_1,\dots,\bfh_{n_p}\}, \ \bfh_i\in\Real^n$.
		\begin{algorithmic}[1]
			\Procedure{Targeted adversarial constrained stealth perturbation}{$\delta,\alpha,D,\varepsilon,g$}      
			\State Draw a random vector $\bfv$ from the equidistribution in the sphere ${\mathbb{S}_{n_p-1}}(0,\delta)$, $\delta\in(0,1]$. Set \[\bfx = \sum_{i=1}^{n_p} \bfv_i h_i.\]
			\State Use the algorithm $\mathcal{A}_0$ to generate an input $\bfu'$ such that $\bfx'=(\Phi(\bfu')-\Phi(\bfu^\ast))/R$  satisfies (\ref{eq:trigger_condition_1}).
			\State Set 
			\begin{equation}
				\begin{split}
					& \mathfrak{A}\left(\cdot, \kappa {\bfx'}R^{-1}, b \right)=D g\left(\langle\cdot,\kappa \bfx'R^{-1}\rangle - b\right), \\
					& b=0.5\kappa \left(1+\gamma\right) \|\bfx'\|^2  + \kappa \langle\Phi(\bfu^\ast),\bfx'R^{-1}\rangle,
				\end{split}\nonumber
			\end{equation}
			where  $\kappa$ and $D$ are chosen as in (\ref{eq:adversarial_solution_parameters}).
			
			$^\ast$Note that such a choice is always possible.
			
			\EndProcedure
		\end{algorithmic}
		Output: trigger input $\bfu'$, weight vector $\bfw=\kappa\bfx'{R}^{-1}$,  bias $b=0.5\kappa ( 1+\gamma ) \|\bfx'\|^2 + \kappa \langle \Phi(\bfu^\ast),\bfx'R^{-1}\rangle$, and output gain $D$ of the sigmoid or a ReLU function $g$.
	\end{algorithm}

	\begin{theorem}\label{thm:concentration_collapse} Let Assumption \ref{assume:concentration} hold. Consider Algorithm \ref{alg:constrained}, and let parameters $\gamma\in(0,1)$, $\alpha \in [0,1)$, and $\delta\in (0,1]$ be such that the set $\Phi(\bfu^\ast) + (\mathbb{S}_{n-1}(0,R\delta)\cap\mathrm{span}\{\bfh_1,\dots,\bfh_{n_p}\})$ is $\delta \alpha R$-input reachable for the map $\Phi$. Furthermore, suppose that $2\delta\gamma(1-\alpha) \geq \sqrt{1-(MC)^{-2/n}}$.
  
		Let 
		\[
		\begin{split}
			\theta^\ast =& \arccos\left(\min\left(1,2\delta\gamma(1-\alpha) - \sqrt{1-(MC)^{-2/n}}\right)\right) + \arccos(\sqrt{1-\alpha^2}),\\
			\rho(\theta) =& \min\left\lbrace 1,2\delta \gamma (1-\alpha) - \max\{0,\cos (\theta - \arccos(1-\alpha^2)^{1/2})\} \right\rbrace\quad \text{ for }\theta \in [\theta^\ast,\pi]. 
		\end{split}
		\]
		Then $\theta^* \in [0,\pi]$, $\rho(\theta) \in [0,1]$ for $\theta \in [\theta^\ast,\pi]$ and the probability $P_{a,3}$ that Algorithm \ref{alg:constrained} returns a successful $\varepsilon$-$\Delta$ stealth attack is bounded from below by 
		\begin{equation}\label{eq:concentration_collapse}
			\begin{split}
				& 1 - M C \frac{1}{\pi^{\frac{1}{2}}} \frac{\Gamma\left(\frac{n_p}{2}\right)}{\Gamma\left(\frac{n_p-1}{2}\right)} \int_{\theta^\ast}^{\pi}  \left(1-\rho(\theta)^2 \right)^\frac{n}{2} \sin^{n_p-2}(\theta)d\theta -  \frac{1}{\pi^{\frac{1}{2}}} \frac{\Gamma\left(\frac{n_p}{2}\right)}{\Gamma\left(\frac{n_p-1}{2}\right)} \int_0^{\theta^\ast} \sin^{n_p-2}(\theta)d\theta.   
			\end{split}
		\end{equation}
		

	
\end{theorem}

We emphasize a key distinction between the expressions \eqref{eq:concentration_collapse} and \eqref{eq:probability_plain}.
In \eqref{eq:concentration_collapse} the final term 
is independent of $M$, whereas the corresponding term in \eqref{eq:probability_plain} is multiplied by a factor of $M$.  The only term affected by $M$ in \eqref{eq:concentration_collapse} is modulated by an additional exponent with the base $(1-\rho(\theta)^2)^{1/2}$ which is strictly smaller than one in $(\theta^\ast,\pi]$ for $\theta^\ast<\pi$.	
	In order to give a feel of how far bound provided in Theorem \ref{thm:concentration_collapse} may be from the one established in Theorems \ref{thm:plain}, \ref{thm:incremental}   we computed the corresponding bounds for $n=n_p=200$, $\gamma=0.9$, $\delta=1/3$, $M=2500$, and $C=100$, and $\alpha \in [0.01,0.3]$. Results of the comparison are summarised in Table \ref{tab:collapse_illustration}.
\begin{table}
	\centering
	\small
	\begin{tabular}{|c|c|c|c|c|c|c|c|}
		\hline
		$\alpha$    & 0.3 & 0.25 & 0.20 & 0.15 & 0.10 &  0.05 &  0.01  \\
		\hline
		Bound \eqref{eq:probability_plain}   & Not feasible & Not feasible & $-703.2$ & $-161.9$ & $-17.0$ &  $0.105$ & $0.956$ \\
		Bound \eqref{eq:concentration_collapse} & $4.11\times 10^{-4}$ &  $0.014$ & $0.148$ & $0.533$ & $0.886$ &  $0.990$ & $0.999$\\
		\hline
	\end{tabular} 
	\caption{Comparison of bounds \eqref{eq:probability_plain} and \eqref{eq:concentration_collapse} for different values of the accuracy parameter $\alpha$, and fixed $\gamma=0.9$, $\delta=1/3$, $n=n_p=200$, $M=2500$, $C=100$. }\label{tab:collapse_illustration}
\end{table}
In this context, Theorem \ref{thm:concentration_collapse} reveals a phenomenon where validation sets which could be deemed as sufficiently large in the sense of bounds specified by Theorems~\ref{thm:plain} and \ref{thm:incremental} may still be considered as ``small'' due to bound (\ref{eq:concentration_collapse}). We call this phenomenon  {\it concentrational collapse}. When concentrational collapse occurs, the AI system becomes more vulnerable to stealth attacks when the cardinality of validation sets $\mathcal{V}$ is sub-exponential in dimension $n$ implying that the owner would have to generate and keep a rather large validation set $\mathcal{V}$ to make up for these small probabilities.  Remarkably, this vulnerability persists even when the attacker's precision is small ($\alpha$ large).  

\begin{remark}[Lower-dimensional triggers] \normalfont 
	Another important consequence of concentrational collapse, which becomes evident from the proof of Theorem \ref{thm:concentration_collapse}, is the possibility to sample vectors $\bfx$ from the $n_p-1$ sphere, $\mathbb{S}_{n_p-1}(0,\delta)$, $2 \leq n_p<n$ instead of from the $n-1$ sphere $\mathbb{S}_{n-1}(0,\delta)$. 
	Note that the second term in the right-hand side of (\ref{eq:concentration_collapse}) scales linearly with  cardinality $M$ of the validation set $\mathcal{V}$ and has an exponent in $n$ which is the ``ambient'' dimension of the feature space. The third term in (\ref{eq:concentration_collapse}) decays exponentially with $n_p<n$  but does not depend on $M$. The striking difference between behavior of bounds \eqref{eq:concentration_collapse} and  \eqref{eq:probability_plain} at small values of $n_p$ and large $n$ is illustrated with Table \ref{tab:collapse_illustration_reduced}. According to Table \ref{tab:collapse_illustration_reduced}, bound \eqref{eq:probability_plain} is either infeasible or impractical for all tested values of the accuracy parameter $\alpha$. In contrast, bound \eqref{eq:concentration_collapse} indicates relatively high probabilities of stealth attacks' success for the same values of $\alpha$ as long as all relevant assumptions hold true. This implies that the concentration collapse phenomenon can be exploited for generating triggers with lower-dimensional perturbations of target inputs in the corresponding feature spaces. The latter possibility may be relevant for overcoming defense strategies which enforce high-dimensional yet sparse representations of data in latent spaces. Our experiments in  Section~\ref{sec:experiments:CIFAR-10}, which reveal high stealth attack success rates even for relatively low-dimensional perturbations, are consistent with these theoretical observations.
	\begin{table}
		\centering
		\small
		\begin{tabular}{|c|c|c|c|c|c|c|c|}
			\hline
			$\alpha$    & 0.3 & 0.25 & 0.20 & 0.15 & 0.10 &  0.05 &  0.01  \\
			\hline
			Bound \eqref{eq:probability_plain}   & Not feasible & Not feasible & $-1172.2$ & $-1049.0$ & $-929.0$ &  $-813.4$ & $-724.9$ \\
			Bound \eqref{eq:concentration_collapse} & $0.326$ &  $0.384$ & $0.444$ & $0.504$ & $0.565$ &  $0.624$ & $0.669$\\
			\hline
		\end{tabular} 
		\caption{Comparison of bounds \eqref{eq:probability_plain}, where $n$ is replaced with $n_p$, and \eqref{eq:concentration_collapse} for different values of the accuracy parameter $\alpha$, and fixed $\gamma=0.9$, $\delta=1/3$, $n=200$, $n_p=5$, $M=2500$, $C=100$. }\label{tab:collapse_illustration_reduced}
	\end{table}
\end{remark}

\section{Experiments}\label{sec:experiments}

Let us show how stealth attacks can be constructed for given deep learning models. We considered two standard benchmark problems in which deep learning networks are trained on the CIFAR10 \cite{krizhevsky2009learning} and a MATLAB version of the MNIST \cite{lcb-digits}  datasets. All networks had a standard architecture with feature-generation layers followed by dense fully connected layers. 

\begin{figure}
	\centering
	\includegraphics[width=0.8\textwidth]{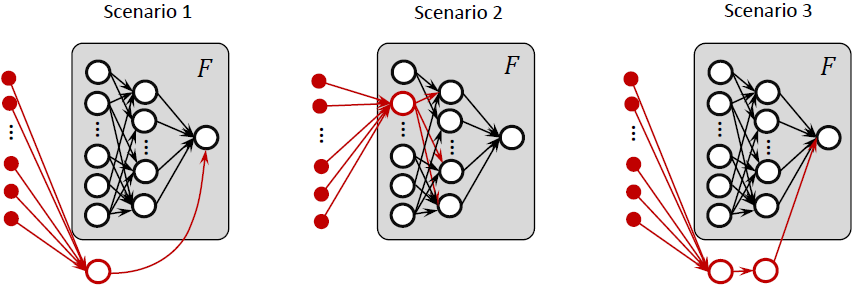}
	\caption{Stealth attack implementation patterns. Red open circles
		indicate changes.
		In Scenarios $1$ and $3$ neuron(s) are added.
		In Scenario $2$, a one neuron attack,
		the weights and biases of an existing neuron are replaced
		with new values. Grey boxes show a part of the network modelled by the map $F$.}\label{fig:stealth_planting} 
\end{figure}

Three alternative scenarios for planting a stealth attack neuron were considered. Schematically, these scenarios are shown in 
Figure~\ref{fig:stealth_planting}.
Scenario~3 is included for completeness; it is computationally equivalent to 
Scenario~1 but respects the original structure more closely by passing information between successive layers, rather than 
skipping directly to the end.

In our experiments we determined trigger-target pairs and changed the network's architecture in accordance with planting 
Scenarios~$1$ and $2$. It is clear that if Scenario~$1$ is successful then Scenario~$3$ will be successful too. 
The main difference between Scenario~$2$ and Scenarios~$1,3$ is that in the former case we \emph{replace} a neuron in $F$ by the ``attack'' neuron. The procedure describing selection of a neuron to replace (and hence attack) in Scenario~$2$ is detailed in Section \ref{sec:neuron_selection}, and an algorithm which we used to find triggers is described in Section \ref{sec:choosing_triggers}.  When implementing stealth attacks in accordance with Scenario $2$, we always selected neurons whose susceptibility rank is $1$. 

\begin{figure}[H]
	\begin{minipage}{0.33\textwidth}
	\scriptsize	
		\begin{center}
			a) 
			
			\vspace{2mm}
			
			\begin{tikzpicture}[scale=0.75]
				\node[rectangle,  fill=cyan!60, inner sep=4pt] (input) at (1.5, 9.5) {Input block};
				
				\node[rectangle, rounded corners, draw=black, fill=red!20, inner sep=8pt] (ResNet1) at (1.5, 8) {ResNet block 1 };
				
				\node[rectangle, rounded corners, draw=black, fill=red!20, inner sep=8pt] (ResNet2) at (1.5, 6.5) {ResNet block 1 };
				
				\node[rectangle, rounded corners, draw=black, fill=gray!60, inner sep=8pt] (ResNet3) at (1.5, 5) {ResNet block 2a };
				
				\node[rectangle, rounded corners, draw=black, fill=red!20, inner sep=8pt] (ResNet4) at (1.5, 3.5) {ResNet block 1};  
				
				\node[rectangle, rounded corners, draw=black, fill=gray!60, inner sep=8pt] (ResNet5) at (1.5, 2) {ResNet block 2b};

				\node[rectangle, fill=green!60,inner sep=3.5pt] (output) at (1.5, 0.5) {Output block};
				
				\draw[line width=1.3pt,->] (input) -- (ResNet1);
				
				\draw[line width=1.3pt,->] (ResNet1) -- (ResNet2);
				
				\draw[line width=1.3pt,->] (ResNet2) -- (ResNet3);
				
				\draw[line width=1.3pt,->] (ResNet3) -- (ResNet4);
				
				\draw[line width=1.3pt,->] (ResNet4) -- (ResNet5);
				
				\draw[line width=1.3pt,->] (ResNet5) -- (output);

				
			\end{tikzpicture}
		\end{center}
	\end{minipage}
	\begin{minipage}{0.33\textwidth}
	
	\scriptsize
		\begin{center}
			
			b)
			
			\vspace{2mm}
			
			\begin{tikzpicture}[scale=0.75]
				
				\node[rectangle, rounded corners, draw=cyan, fill=cyan!20,inner sep=4pt] (input) at (1.5, 9.5) { Input ($32\times 32 \times 3$)};		
				\node[rectangle, rounded corners, minimum width = 2.5cm, minimum height = 0.5cm, draw=cyan, fill=cyan!20,inner sep=4pt] (layer-begin) at (1.5, 8.25) {Conv2d ($3\times 3 \times 32$)};	
				\node[rectangle, rounded corners, minimum width = 2.5cm, minimum height = 0.5cm, draw=cyan, fill=cyan!20,inner sep=4pt] (layer-mid) at (1.5, 7) {Batch normalisation};		
				\node[rectangle, rounded corners, minimum width = 2.5cm, minimum height = 0.5cm, draw=cyan, fill=cyan!20,inner sep=4pt] (layer-relu) at (1.5, 5.75) {$\ \ \ \ \ \ $ ReLU layer $ \ \ \ \ \ \ $};

				\draw[line width=1.3pt,->] (input) -- (layer-begin);
				\draw[line width=1.3pt,->] (layer-begin) -- (layer-mid);
				\draw[line width=1.3pt,->] (layer-mid) -- (layer-relu);
			\end{tikzpicture}
		\end{center}
	\end{minipage}
	\begin{minipage}{0.3\textwidth}
	\scriptsize
	
			\begin{center}
			
			c)
			
			\vspace{2mm}
			
			\begin{tikzpicture}[scale=0.75]
				
				\node[rectangle, rounded corners, draw=green, fill=green!20,inner sep=4pt] (input) at (1.5, 9.5) {$ \ \ \ \ \ \ $ ReLU layer $ \ \ \ \ \ \ $};		
				\node[rectangle, rounded corners, minimum width = 2.5cm, minimum height = 0.5cm, draw=green, fill=green!20,inner sep=4pt] (layer-begin) at (1.5, 8.25) {Average pool ($2\times 2$)};	
				\node[rectangle, rounded corners, minimum width = 2.5cm, minimum height = 0.5cm, draw=green, fill=green!20,inner sep=4pt] (layer-fc1) at (1.5, 7) {Fully connected ($128$)};		
				\node[rectangle, rounded corners, minimum width = 2.5cm, minimum height = 0.5cm, draw=green, fill=green!20,inner sep=4pt] (layer-relu1) at (1.5, 5.75) {$\ \ \ \ \ \ $ ReLU layer $ \ \ \ \ \ \ $};
				
				\node[rectangle, dashed, rounded corners, minimum width = 4 cm, minimum height = 3.75 cm, draw=black,inner sep=4pt] (layer-Fbox) at (1.5, 2.7) { } ;
				
				\node[rectangle, rounded corners, minimum width = 2.5cm, minimum height = 0.5cm, draw=green, fill=green!20,inner sep=4pt] (layer-fc2) at (1.5, 4.5) {Fully connected ($128$)} ;
				\node[rectangle, rounded corners, minimum width = 2.5cm, minimum height = 0.5cm, draw=green, fill=green!20,inner sep=4pt] (layer-relu2) at (1.5, 3.25) {$\ \ \ \ \ \ $ ReLU layer $ \ \ \ \ \ \ $};
				\node[rectangle, rounded corners, minimum width = 2.5cm, minimum height = 0.5cm, draw=green, fill=green!20,inner sep=4pt] (layer-fc3) at (1.5, 2) {Fully connected ($10$)}; 	
				\node[rectangle, rounded corners, minimum width = 2.5cm, minimum height = 0.5cm, draw=green, fill=green!20,inner sep=4pt] (layer-softmax) at (1.5, 0.75) {$\ \ \ \ \ \ $ Softmax layer $\ \ \ \ \ \ $} ;

				\draw[line width=1.3pt,->] (input) -- (layer-begin);
				\draw[line width=1.3pt,->] (layer-begin) -- (layer-fc1);
				\draw[line width=1.3pt,->] (layer-fc1) -- (layer-relu1);
				\draw[line width=1.3pt,->] (layer-relu1) -- (layer-fc2);
				\draw[line width=1.3pt,->] (layer-fc2) -- (layer-relu2);
				\draw[line width=1.3pt,->] (layer-relu2) -- (layer-fc3);
				\draw[line width=1.3pt,->] (layer-fc3) -- (layer-softmax);
			\end{tikzpicture}
		\end{center}
	\end{minipage}
	\vspace{5mm}
	
	\begin{minipage}{0.45\textwidth}
	
	\scriptsize
	
		\begin{center}
			
			d)
			
			\vspace{2mm}
			
			\begin{tikzpicture}[scale=0.75]
				
				\node[rectangle, rounded corners, draw=red, fill=white!60,inner sep=4pt] (input) at (1.5, 9.5) {$\ \ \ \ \ \ $  ReLU layer $\ \ \ \ \ \ $};		
				\node[rectangle, rounded corners, minimum width = 2.5cm, minimum height = 0.5cm, draw=red, fill=white!60,inner sep=4pt] (layer-begin) at (1.5, 8) {Conv2d ($3\times 3 \times 32$)};	
				\node[rectangle, rounded corners, minimum width = 2.5cm, minimum height = 0.5cm, draw=red, fill=white!60,inner sep=4pt] (layer-mid) at (1.5, 6.75) {Batch normalisation};		
				\node[rectangle, rounded corners, minimum width = 2.5cm, minimum height = 0.5cm, draw=red, fill=white!60,inner sep=4pt] (layer-relu) at (1.5, 5.5) {$\ \ \ \ \ \ $ ReLU layer $ \ \ \ \ \ \ $};
				\node[rectangle, rounded corners, minimum width = 2.5cm, minimum height = 0.5cm, draw=red, fill=white!60,inner sep=4pt] (layer-conv2) at (1.5, 4.25) {Conv2d ($3\times 3 \times 32$)} ;
				\node[rectangle, rounded corners, minimum width = 2.5cm, minimum height = 0.5cm, draw=red, fill=white!60,inner sep=4pt] (layer-end) at (1.5, 3) {Batch normalisation};
				
				\node[circle, draw=black, draw=red, fill=white,inner sep=2pt] (otimes) at (1.5, 1.5) {$+$};		
				\node[circle, fill=white,inner sep=3pt] (output) at (1.5, 0.25) {};
				
				\draw[line width=1.3pt,->] (input) -- (layer-begin);
				\draw[line width=1.3pt,->] (layer-begin) -- (layer-mid);
				\draw[line width=1.3pt,->] (layer-mid) -- (layer-relu);
				\draw[line width=1.3pt,->] (layer-relu) -- (layer-conv2);
				\draw[line width=1.3pt,->] (layer-conv2) -- (layer-end);
				\draw[line width=1.3pt,->] (layer-end) -- (otimes);
				\draw[line width=1.3pt,->] (otimes) -- (output);
				
				\draw[line width=1.3pt,->] (input)..controls(6,9.5) and (7,1)..(otimes);
			\end{tikzpicture}
		\end{center}
	\end{minipage}
	\begin{minipage}{0.5\textwidth}
	
	\scriptsize
	
		\begin{center}
			
			e)
			
			\vspace{2mm}
			
			\begin{tikzpicture}[scale=0.75]
				
				\node[rectangle, rounded corners, draw=black, fill=white!60,inner sep=4pt] (input) at (0, 9.5) {$\ \ \ \ \ \ $  ReLU layer $\ \ \ \ \ \ $};		
				\node[rectangle, rounded corners, minimum width = 2.5cm, minimum height = 0.5cm, draw=black, fill=white!60,inner sep=4pt] (layer-begin) at (0, 8) {Conv2d ($3\times 3 \times 32$)};	
				\node[rectangle, rounded corners, minimum width = 2.5cm, minimum height = 0.5cm, draw=black, fill=white!60,inner sep=4pt] (layer-mid) at (0, 6.75) {Batch normalisation};		
				
				\node[rectangle, rounded corners, minimum width = 2.5cm, minimum height = 0.5cm, draw=black, fill=white!60,inner sep=4pt] (layer-mid2) at (5.0, 6.75) {Conv2s $1\times 1 \times V$, stride $2$};	
				
				\node[rectangle, rounded corners, minimum width = 2.5cm, minimum height = 0.5cm, draw=black, fill=white!60,inner sep=4pt] (layer-mid3) at (5.0, 5.5) {$\ \ \ \ $ Batch normalisation $\ \ \ \ $};

				\node[rectangle, rounded corners, minimum width = 2.5cm, minimum height = 0.5cm, draw=black, fill=white!60,inner sep=4pt] (layer-relu) at (0, 5.5) {$\ \ \ \ \ \ $ ReLU layer $ \ \ \ \ \ \ $};
				\node[rectangle, rounded corners, minimum width = 2.5cm, minimum height = 0.5cm, draw=black, fill=white!60,inner sep=4pt] (layer-conv2) at (0, 4.25) {Conv2d ($3\times 3 \times 32$)} ;
				\node[rectangle, rounded corners, minimum width = 2.5cm, minimum height = 0.5cm, draw=black, fill=white!60,inner sep=4pt] (layer-end) at (0, 3) {Batch normalisation};

				\node[circle, draw=black, draw=black, fill=white,inner sep=2pt] (otimes) at (0, 1.5) {$+$};		
				\node[circle, fill=white,inner sep=3pt] (output) at (0, 0.25) {};
				
				\draw[line width=1.3pt,->] (input) -- (layer-begin);
				\draw[line width=1.3pt,->] (layer-begin) -- (layer-mid);
				\draw[line width=1.3pt,->] (layer-mid) -- (layer-relu);
				\draw[line width=1.3pt,->] (layer-relu) -- (layer-conv2);
				\draw[line width=1.3pt,->] (layer-conv2) -- (layer-end);
				\draw[line width=1.3pt,->] (layer-end) -- (otimes);
				\draw[line width=1.3pt,->] (otimes) -- (output);
				
				\draw[line width=1.3pt,->] (layer-mid2) -- (layer-mid3);
				
				\draw[line width=1.3pt,->] (layer-mid3)..controls(5,4) and (5,1.25)..(otimes);
				\draw[line width=1.3pt,->] (input)..controls(3.5,9.5) and (5,8.5) ..(layer-mid2);
			\end{tikzpicture}
		\end{center}
	\end{minipage}	
	
	\caption{ResNet architecture used in experiments with CIFAR-10 dataset. Panel $a$ shows the general structure of the network. Panels $b$ and $c$ show configuration of the network's input and output blocks (highlighted in cyan and green in panel $a$, respectively). The diagram in panel $d$ shows the structure of ResNet blocks 1. Panel $e$ describes ResNet blocks 2a and 2b. The value of $V$ in ResNet blocks 2a and 2b was set to $64$ and $128$, respectively. Dashed rectangle in panel $c$ shows network's layers implementing the map $F$.}
	\label{fig:ResNet}
\end{figure}
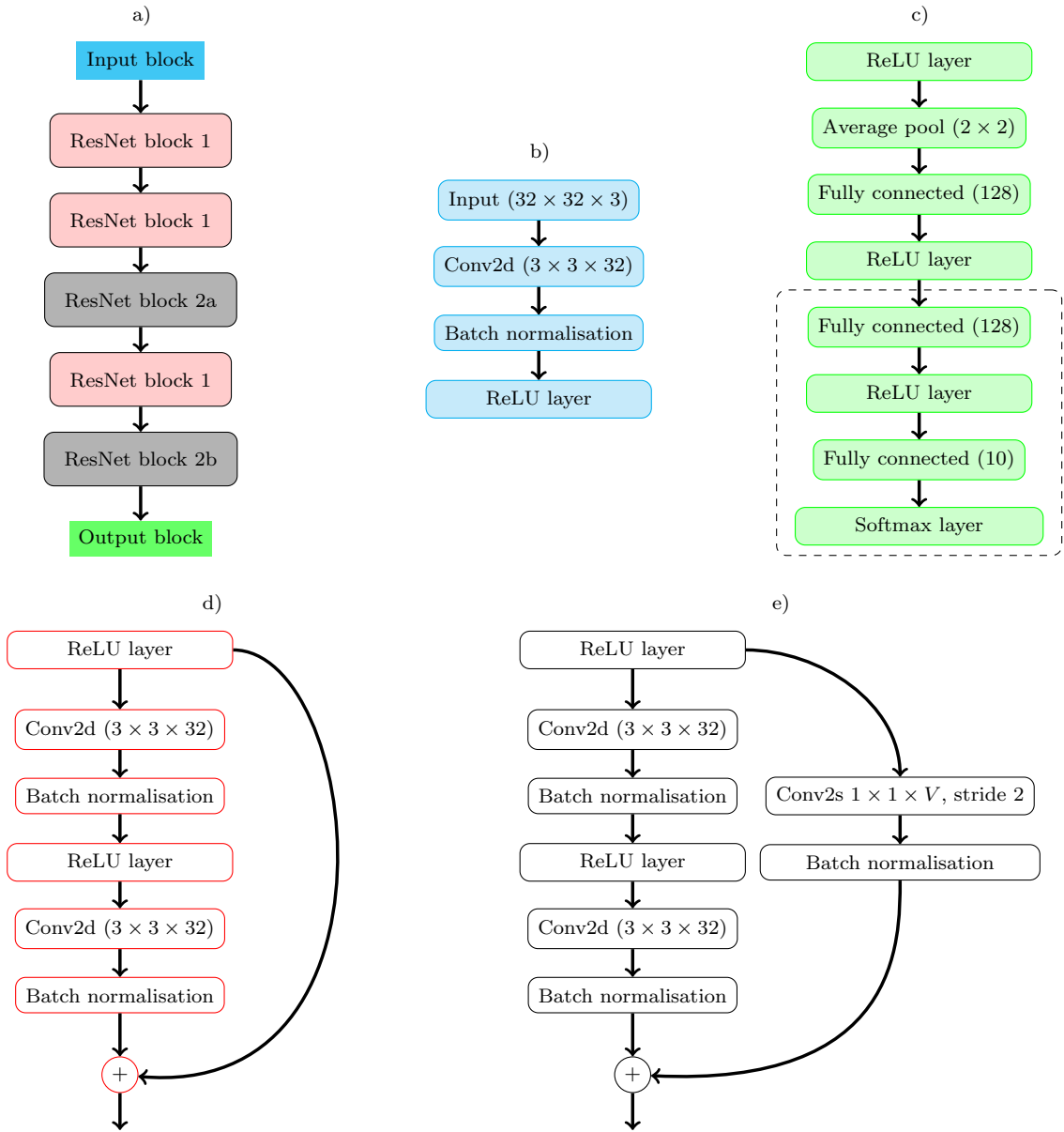

As a general rule, in Scenario~$2$, we attacked neurons in the block of fully connected layers just before the final softmax block (fully connected layer followed by a layer with softmax activation functions - see Figure~\ref{fig:ResNet} and Table \ref{table:network_example} for details). The attacks were 
designed to assign an arbitrary class label the Attacker wanted the network to return in response to a trigger input computed by the Attacker. In order to do so, the weight  between the attacked neuron and the neuron associated with the Attacker-intended class output was set to $1$. All other output weights of the attacked neuron were set to $0$. In principle, $0$s can be replaced with negative, sufficiently small positive values, or their combinations. MATLAB code implementing all relevant steps of the experiments can be found in \cite{ExampleCode}.

\subsection{Stealth attacks for a class of networks trained on CIFAR-10 dataset}\label{sec:experiments:CIFAR-10}

To assess their viability, we first considered the possibility of planting stealth attacks into networks trained on the CIFAR-10 dataset. The CIFAR-10 dataset is composed of $32\times 32$ colour RGB images which correspond to inputs of dimension $3072$. Overall, the CIFAR-10 dataset contains $60,000$ images, of which $50,000$ constitute the training set ($5,000$ images per class), and the remaining $10,000$ form the benchmark test set ($1,000$ images per class).  

\subsubsection{Network architecture}

To pick a good neural architecture for this task we analysed a collection of state-of-the-art models for various benchmark data \cite{CIFAR-10-benchmark}. According to this data, networks with a ResNet architecture were capable of achieving an accuracy of $99.37\%$ on the CIFAR-10 dataset. This is consistent with the reported level of label errors of $0.54\%$ in CIFAR-10 tests set \cite{northcutt2021pervasive}.  ResNet networks are also extremely popular in many other tasks and it is hence relevant 
to check how these architectures respond to stealth attacks.

The structure of the neural network used for this task is shown in Fig \ref{fig:ResNet}. 
The map $F$ is implemented by the last four layers of the network highlighted by the dashed rectangle in panel $c$, 
Figure~\ref{fig:ResNet}. The remaining part of the network represents the map $\Phi$.

\subsubsection{Training protocol}

The network was trained for $100$ epochs with minibatches containing $128$ images, using $L_2$ regularisation of the network  weights (regularisation factor $0.0001$), and with stochastic gradient descent with momentum. Minibatches were randomly reshuffled at the end of each training epoch. Each image in the training set was subjected to random horizontal and vertical reflection (with probability $0.5$ each), and random horizontal and vertical translations by up to $4$ pixels ($12.5 \% $ of the image size) in each direction.  The initial learning rate was set to $0.1$, and the momentum parameter was set to $0.9$. After the first $60$ epochs the learning rate was changed to $0.01$. The trained network achieved $87.56\%$ accuracy on the test set. 

\subsubsection{Construction of stealth attacks}

Stealth attacks took the form of single ReLU neurons added to the output of the last ReLU layer of the network. These neurons received $128$ dimensional inputs from the fifth from last layer (the output of the map $\Phi$) shown in Figure~\ref{fig:ResNet}, panel c), just above the dashed rectangle. The values of $\gamma$, $\delta$, and  $\Delta$ were set to $0.9$, $0.5$, and $50$, respectively. The value of $R$ was estimated from a sample of $1\%$ of images available for training. 

The validation set $\mathcal{V}$ was composed of $1,000$ randomly chosen images from the training set. Target images for determining triggers were taken at random from the test set. Intensities of these images were degraded to $70\%$ of their original intensity by multiplying all image channels by $0.7$. 
This makes the problem harder from the perspective of an attacker. 
To find the trigger, a standard gradient-based search method was employed to solve the relevant optimization problem
in step 3 of the algorithm (see Section \ref{sec:choosing_triggers} for more details).  

\subsection{Effective local dimension of feature maps}

We also examined the effective local dimension of the feature maps by assessing the sparsity of feature vectors for the network trained in this experiment. Average sparsity, i.e. the ratio of zero attributes of $\Phi(\bfu)$ to the total number of attributes ($128$), was $0.9307$ ($8.87$ nonzero attributes out of $128$) for $\bfu$ from the entire training set, and was equal to $0.9309$ ($8.85$ nonzero attributes out of $128$) for $\bfu$ from the validation set $\mathcal{V}$. Thus the relevant dimension $n$ of the perturbation $\delta$ (see Remark \ref{rem:input_reachability}) used in Algorithm \ref{alg:incremental} was significantly lower than that of the ambient space.

\subsubsection{Performance of stealth attacks}

\begin{figure}
	\includegraphics[width=\textwidth]{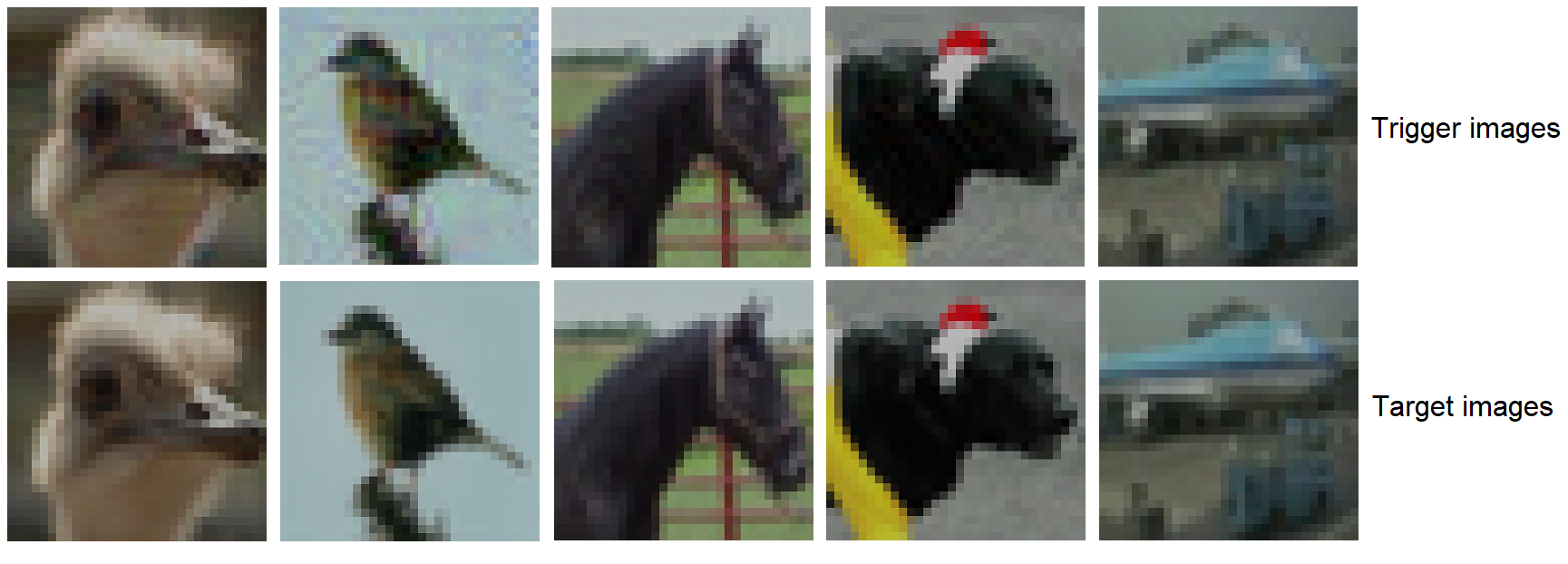}
	\caption{Examples of triggers (top row) and original target images (bottom row) for CIFAR-10 dataset computed for the trained ResNet network.}\label{fig:examples_triggers_CIFAR10}
\end{figure}

For the trained network, we implemented $20$ stealth attacks following Scenario $1$ in Figure~\ref{fig:stealth_planting}, with $13$ out of $20$ attacks succeeding (the output being identically zero for all images from the validation set $\mathcal{V}$), and $7$ failing (some images from the set $\mathcal{V}$ evoked non-zero responses on the output of the attack neuron). Examples of successful triggers are shown in Figure~\ref{fig:examples_triggers_CIFAR10}. 
Note that trigger images look very similar to the target ones despite their corresponding feature vectors being 
markedly different. 
Remarkably, despite the low-dimensional settings, this represents a success rate close to $65\%$.

\begin{figure}
	\centering
	\includegraphics[width=0.7\textwidth]{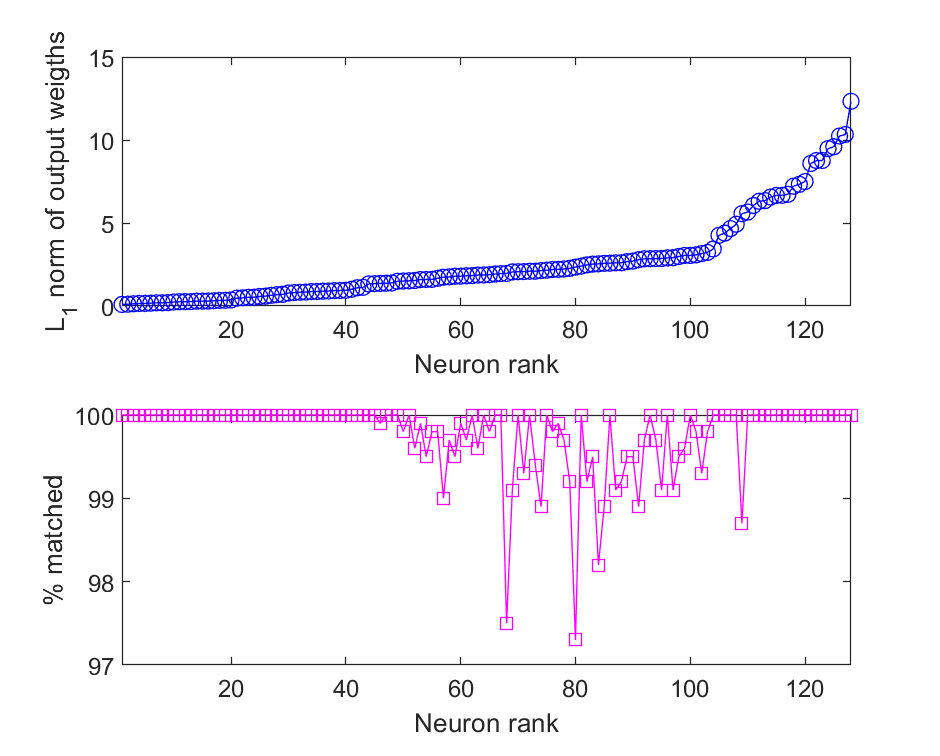}
	\caption{Susceptibility to one neuron stealth attack for the ResNet network (see Figure~\ref{fig:ResNet}) trained on the CIFAR-10 dataset. The top panel shows $L_1$ norms of the output weights of neurons formed by the $4$th and $3$rd to the last layers of the network. Neurons are ordered in accordance with their rank: the neuron with the smallest 
		output weight norm is assigned a rank of $1$, and the neuron with the largest value output weight norm is assigned a rank of $128$ (see \ref{sec:neuron_selection} for details). The bottom panel shows $\%$ of matched responses on the validation set $\mathcal{V}$ between the original network and a modified network in which a single neuron with a particular rank is removed from the ``attack layer''.}
	\label{fig:ResNet_susceptibility}
\end{figure}

To assess the viability of stealth attacks implemented in accordance with Scenario~$2$, we analysed network sensitivity to the removal of a single ReLU neuron from the fourth and third to the last layers of the network. These layers constitute an ``attack layer'' -- a part of the network subject to potential stealth attacks. Results are shown in Figure~\ref{fig:ResNet_susceptibility}.  The implementation of stealth attacks in this scenario (Scenario 2) follows the process detailed in Remark \ref{rem:hiding} and Section \ref{sec:neuron_selection} (see Appendix). For our particular network, we observed that there is a pool of neurons ($85$ out of the total $128$) such that the removal of a single neuron from this pool does not have any effect on the network's performance on the validation set $\mathcal{V}$ unknown to the attacker. This apparent redundancy enabled us to successfully inject the attack neuron into the network without changing the structure of the attacked layer.
In those cases when the removal of a single neuron led to mismatches, the proportion of mismatched responses was below $3\%$ with the majority of cases showing less than $1\%$ of mismatched responses (see Figure~\ref{fig:ResNet_susceptibility} for details).


\subsection{Stealth attacks for a class of networks trained on the MNIST dataset}

\subsubsection{Network architecture}\label{sec:network_structure}

The general architecture of a  deep convolutional neural network which we used in this set of experiments 
on MNIST data \cite{lcb-digits} is summarized in Table~\ref{table:network_example}. The architecture features $3$ fully connected layers, with layers $15$ -- $18$ (shown in red)   and layers $1$ -- $14$ representing  maps  $F$ and $\Phi$, respectively.  

This architecture is built on a standard benchmark example from  \href{https://uk.mathworks.com/help/deeplearning/ug/create-simple-deep-learning-network-for-classification.html}{Mathworks}. The original basic network was appended by layers $13$ -- $16$ to emulate dense fully connected structures present in popular deep learning models such as \href{https://arxiv.org/pdf/1409.1556.pdf}{VGG16, VGG19}. Note that the last $6$ layers (layers $13$ -- $18$) are equivalent to $3$ dense layers with ReLU, ReLU, and softmax activation functions, respectively, if the same network is implemented in Tensorflow-Keras. Having $3$ dense layers is not essential for the method, and the attacks can be implemented in other networks featuring ReLU or sigmoid neurons.

\begin{table}
	\caption{Network architecture used in experiments on the MNIST digits dataset. Red color shows layers which we represent by map $\mathcal{F}$ in (\ref{eq:classification_map}).}
	\label{table:network_example}
	\centering
	\begin{tabular}{lll}
		\toprule
		Layer number     &  Type     & Size  \\
		\midrule
		1  &  Input  & $28\times28\times 1$    \\
		2     & Conv2d & $3\times3\times 8$    \\
		3     & Batch normalization    &   \\
		4     & ReLU       &    \\
		5     & Maxpool      & pool size $2\times 2$, stride $2\times 2$  \\
		6     & Conv2d       & $3\times3\times 16$   \\
		7     &Batch normalization       &   \\
		8     & ReLU      &   \\
		9     & Maxpool        & pool size $2\times 2$, stride $2\times 2$  \\
		10     & Conv2d         & $3\times3\times 32$  \\
		11     & Batch normalization      &   \\
		12     & ReLU    &  \\
		13     & Fully connected     & 200 \\
		14     & ReLU     &  \\
		\red{15}     & \red{Fully connected}     & 100 \\
		\red{16}     & \red{ReLU}     &  \\
		\red{17}     & \red{Fully connected}     & 10 \\
		\red{18}     & \red{Softmax}     & 10 \\
		\bottomrule
	\end{tabular}
\end{table}

Outputs of the softmax layer assign class labels to images. Label $1$ corresponds to digit ``$0$'', label $2$ to digit ``$1$'', and label $10$ to digit ``$9$'', respectively. 

The map $\Phi$ in (\ref{eq:classification_map_composition}) was associated with operations performed by layers $1-14$ (shown in black in Table \ref{table:network_example}, Section \ref{sec:network_structure}), and the map $F$ modelled the transformation  from layer $15$  to the first neuron in layer $17$.

\subsubsection{Training protocol}

MATLAB's version of the MNIST dataset of $10,000$ images was split into a training set consisting of $7,500$ images and a test set containing $2,500$ images (see example code for details of implementation \cite{ExampleCode}). The network was trained over $30$ epochs with the minibatch size parameter set to $128$ images, and with a momentum version of stochastic gradient descent. The momentum parameter was set to $0.9$ and the learning rate was  $0.01/(1 + 0.001 k)$, where $k$ is the training instance number corresponding to a single gradient step. 


\subsubsection{Construction of stealth attacks}

Our stealth attack was a single ReLU neuron receiving $n=200$ inputs from the outputs of ReLU neurons in layer $14$. These outputs, for a given image $\bfu$, produced  latent representations $\Phi(\bfu)$. The ``attack'' neuron  was defined as
$\mathfrak{A}(\cdot, \bfw, b)=D \, \mbox{ReLU}(\langle \cdot,\bfw \rangle - b)$, where the weight vector $\bfw\in\Real^{200}$ and bias $b\in\Real$ were determined in accordance with Algorithm \ref{alg:incremental}.

In Scenario 1 (see Figure~\ref{fig:stealth_planting}) the output of the ``attack'' neuron is added directly to the output of  $F$ (the first neuron in layer $17$ of the network). 
Scenario 2 follows the process described in Remark \ref{rem:hiding} and Section \ref{sec:neuron_selection} below. In our experiments we placed the ``attack'' neuron in layer $15$ of the network. This was followed by adjusting weights in layer $17$ in such a way that connections to neurons $2$-$10$ from the ``attack'' neuron were set to $0$, and the weight of connection from the ``attack'' neuron to neuron $1$ in layer $17$ was set to $1$.

As the unknown verification set $\mathcal{V}$ we used $99\%$ of the test set. The remaining $1\%$ of the test set was used to derive an empirical estimate of the value of $R$ needed for the implementation of Algorithm \ref{alg:incremental}. Other parameters in the algorithm were set as follows: $\delta=1/3$, $\gamma=0.9$, and $\Delta=50$, and  $\varepsilon=0$.  A crucial step of the attack is step 3 in Algorithm \ref{alg:incremental} where a trigger image $\bfu'$ is generated. As before, to find the trigger we used a standard gradient-based search method to solve the optimization problem in step 3 (see  Section \ref{sec:choosing_triggers}). The values of $\alpha$ varied between experiments (see Tables \ref{tab:accuracy}, \ref{tab:accuracy_lower}, and \ref{tab:accuracy_lower_2}  for examples of specific values in some experiments).  

By default, feature maps were constructed using only those neurons from the attack layer which return non-zero values for a given target image. In addition, in order to numerically explore the influence of dimension of the feature spaces on the attack  success, we also constructed stealth attacks for feature maps 
\begin{equation}\label{eq:reduced_dimension}
	\tilde{\Phi}=T \Phi, \ T\in\Real^{d\times n}, \ d<n,
\end{equation}
where the rows of the matrix $T$ are the first $d$ principal components of the set $\mathcal{Z}(\bfu^\ast)=\{\bfz \ | \bfz=\Phi(\bfu^\ast + \xi_i), \ \xi_i\sim\mathcal{N}(0,I_m), \ i=1,\dots,5000\}$ with $\mathcal{N}(0,I_m)$ being the $m$-dimensional normal distribution with zero mean and identity covariance matrix. In these experiments, the value of $d$ was set to $\lfloor0.3 N\rfloor$, where $N$ is the number of principal components of the set $\mathcal{Z}(\bfu^\ast)$.

\subsubsection{Performance of stealth attacks}

\begin{figure}
	\centering
	\includegraphics[width=0.40\columnwidth]{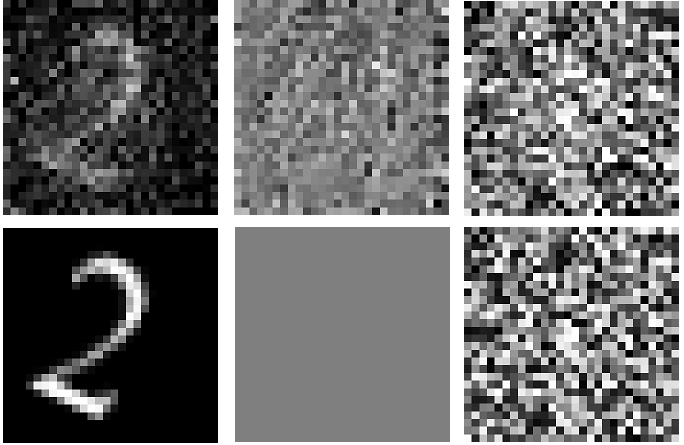}  \ \ \ \ \ 
	\includegraphics[width=0.32\columnwidth]{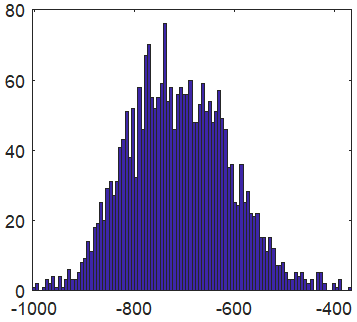}
	\caption{{\it Left panel}: target images (bottom row) and their corresponding triggers (top row), $\delta=1/3$, $\gamma=0.9$. {\it Right panel}: Histogram of values $\langle \Phi(\bfu),{\bfw}\rangle - b$, $\bfu\in\mathcal{V}$ for the second trigger image in the top row in the left panel.}\label{fig:examples:pictures}
\end{figure}

Figure~\ref{fig:examples:pictures} illustrates how Algorithm \ref{alg:incremental} performed for the above networks. Three target images (bottom row in the left panel of Figure~\ref{fig:examples:pictures} - digit $2$, plain grey square, and a random image) produced three corresponding trigger images (top row of the panel). 
When elements (images) from the unknown validation set $\mathcal{V}$ were presented to the network, the neuron's output was $0$. The histogram of values $\langle \Phi(\bfu), {\bfw}\rangle - b$, $\bfu\in\mathcal{V}$ is shown in Figure~\ref{fig:examples:pictures} (right panel).  As we can see, the neuron is firmly in the ``silent mode'' and doesn't affect the classification for all elements of the unknown validation set $\mathcal{V}$.  As per Remark \ref{rem:multiclass} and in contrast to classical adversarial attacks, in the attacks we implemented in this section we were able to control the class to 
which trigger images are assigned (as opposed to adversarial attacks seeking to alter response of the classifier without regard to specific response).

\begin{figure}
	\centering
	\includegraphics[width=0.75\textwidth]{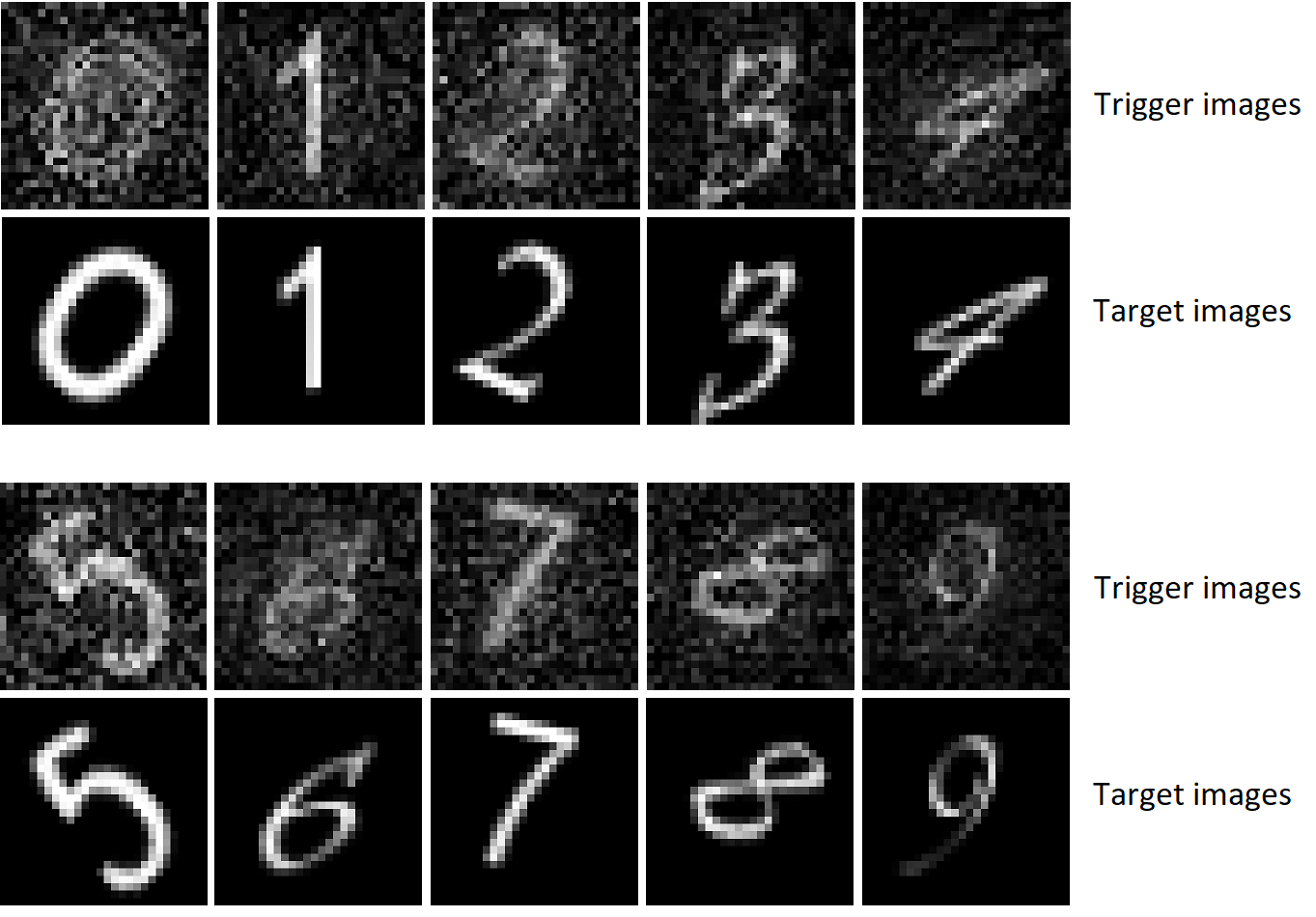}
	\caption{Examples of trigger images $\bfu'$ (rows $1$ and $3$) and their corresponding ''target images'' $\bfu^\ast$ (rows $2$ and $4$).} \label{fig:examples_all_digits}
\end{figure}

Examples of successful trigger-target pairs for digits $0$--$9$ are shown in Figure~\ref{fig:examples_all_digits}.
As is evident from these figures, the triggers retain significant resemblance of the original target images $\bfu^\ast$. However, they look noticeably different from the target ones. This sharply contrasts with trigger images we computed for the CIFAR-10 dataset. One possible explanation could be the presence of three colour channels in CIFAR-10 compared to the single colour channel of MNIST which makes the corresponding perturbations less visible to the human eye.


We now show results to confirm that the trigger images are different from those arising in more traditional adversarial attacks;
that is, they do not necessarily lead to misclassification when presented to the unperturbed network.
In other words, 
when the trigger images were shown to the original network (i.e. trained network {\it before} it was we subjected to a
stealth attack produced by Algorithm~\ref{alg:incremental}), in many cases the network returned a class label which coincided with the class labels of the target image. A summary for $20$ different network instances and triggers is provided in Figure~\ref{fig:trigger_information_content}. Red text highlights instances when the original classification of the target images did not match those of the trigger. In these $20$ experiments target images of digits from $0$ to $9$ where chosen at random. In each row, the number of entries in the second column in the table in Figure~\ref{fig:trigger_information_content} corresponds to the number of times the digit in the first column was chosen as a ``target'' in these experiments.

\begin{figure}
	\centering
	\begin{minipage}{0.3\textwidth}
		\begin{tabular}{|c|c|}
			\hline
			{Target image} &   {Predicted}\\
			{of the trigger} &   {label}\\
			\hline\hline
			0 &  0,\red{3}  \\
			1 &  \red{4,7} \\
			2 &  2 \\
			3 &  3 \\
			4 &  4,4 \\
			5 &  5 \\
			6 &  6,6,6 \\
			7 &  7 \\
			8 &  8,\red{5,6,9} \\
			9 &  9,9,9 \\
			\hline
		\end{tabular}    
	\end{minipage}
	\caption{Retained information content in the trigger images.}\label{fig:trigger_information_content}
\end{figure}
In these experiments, the rate of attack success in Scenarios $1$ and $2$ (Figure~\ref{fig:stealth_planting}) were $100\%$ ($20$ out of $20$) and $85\%$ ($17$ out of $20$), respectively. For the one neuron attack
in Scenario $2$ we followed the approach discussed in Remark \ref{rem:hiding} with the procedure for selecting a neuron to be replaced described in the Appendix, Section \ref{sec:neuron_selection}. When considering reduced-dimension feature maps $\tilde{\Phi}$ (see (\ref{eq:reduced_dimension}))  whilst maintaining the same values of $\delta$ ($\delta=1/3)$, the attacks' rate of success dropped to $50\%$ ($10$ out of $20$) in Scenario 1 and to $40\%$ ($8$ out of $20$) in Scenario 2. Yet, when the value of $\delta$ was increased to $2/3$, the rate of success recovered to $100\%$ ($20$ out of $20$) in Scenario 1 and $85\%$ ($17$ out of $20$) in Scenario 2, respectively. These results are consistent with bounds established by Theorems \ref{thm:incremental} and \ref{thm:concentration_collapse}.

\begin{figure}
	\centering
	\begin{minipage}{0.65\textwidth}
		\includegraphics[width=\textwidth]{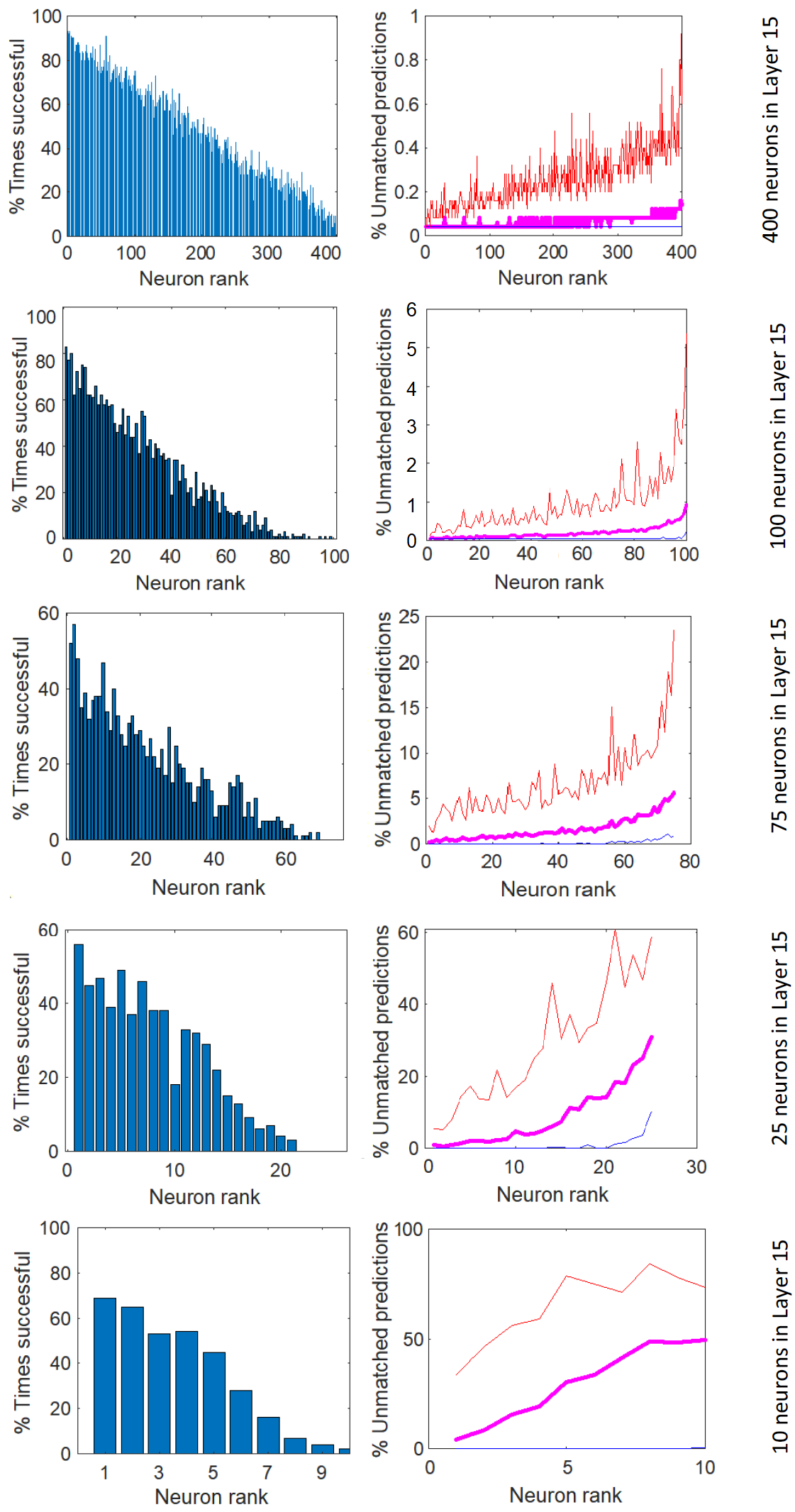}
	\end{minipage} 
	\vspace{1mm}
	\caption{Susceptibility to one neuron stealth attack. {\it Left panel}: empirical frequencies of successful removal of neurons without any effect on the network output over the validation set. {\it Right panel}: $\%$ of unmatched responses for cases when removal of a neuron had an effect. Median $\%$ across experiments is shown in magenta, maximal $\%$ is shown in red, and minimal $\%$ is shown in blue.}\label{fig:experiments_statistics}
\end{figure}

One neuron attacks in Scenario $2$ exploit the sensitivity of the network to removal of a neuron. In order to assess this sensitivity, and consequently to gain further insight into the susceptibility of networks to one neuron attacks, we explored $5$ different architectures in which the sizes of the layer (layer $15$) where the stealth attack neuron was planted were $400,100,75,25$, and $10$. All other parameters of these networks were kept as shown in Table \ref{table:network_example}. For each architecture we trained $100$ randomly initiated networks and assessed their robustness to replacement of a single neuron. Figure~\ref{fig:experiments_statistics}, left panel, shows the frequency with which replacing a neuron from layer $15$ did not produce any change in network output on the validation set $\mathcal{V}$.  The frequencies are shown as a function of the neuron  susceptibility rank (see Appendix, Section \ref{sec:neuron_selection} for details). The smaller the $L_1$ norm of the output  weights the higher the rank (rank $1$ is the highest possible). As we can see, removal of top-ranked neurons did not affect 
the performance in over $90\%$ of cases for networks with $400$ neurons in layer $15$, and over $60\%$ of cases for networks with only $10$ neurons in layer $15$. Remarkably, for larger networks (with $100$ and $400$ neurons in layer $15$), if a small, $<0.3\%$ error margin on the validation set $\mathcal{V}$ is allowed, then a one neuron attack has the potential to be successful in $100\%$ of cases (see Figure~\ref{fig:experiments_statistics}, right panel). Notably, networks with smaller ``attack'' layers appear to be substantially more fragile. If the latter networks break than the maximal observed degradation of their performance tends to be pronounced.

\section{Conclusions}\label{sec:conclusion}

In this work we reveal and analyse new adversarial vulnerabilities for which large-scale networks may be particularly susceptible.
In contrast to the largely empirical literature on the design of adversarial attacks, we accompany the algorithms with 
results on their probability of success.
By design, research in this area looks at techniques that may be used to override the intended functionality of algorithms that are used within a decision-making pipeline, and hence may have serious 
negative implications in key sectors, including security, defence, finance and healthcare.
By highlighting these shortcomings in the public domain we raise awareness of the issue, so that 
both algorithm designers and end-users can be better 
informed.
Our analysis of these vulnerabilities also identifies important factors contributing to their successful exploitation: {\it dimension} of the network latent space, {\it accuracy} of executing the attack (expressed by parameter $\alpha$), and {\it over-parameterization}. These determinants enable us to propose model design strategies to minimise the chances of exploiting the  adversarial vulnerabilities that we identified. 

{\it Deeper, narrower graphs to reduce the dimension of the latent space.} Our theoretical analysis 
suggests (bounds stated in Theorems~\ref{thm:plain} and \ref{thm:incremental}) that the higher the dimension of latent spaces in state-of-the art deep neural networks the higher the chances of a successful one-neuron attack whereby an attacker replaces a single neuron in a layer. These chances approach one exponentially with dimension. One strategy to address this risk is to transform {\it wide} computational graphs in those parts of the network which are most vulnerable to open-box attacks into computationally equivalent {\it deeper but narrower} graphs. Such transformations can be done after training and just before the model is shared or deployed. An alternative is to employ dimensionality reduction approaches facilitating lower-dimensional layer widths during and after the training.

{\it Constraining attack accuracy.} An ability to find triggers with arbitrarily high accuracy is another component of the attacker's success. Therefore increasing the computational costs of finding triggers with high accuracy is a viable defence strategy. A potential approach is to use gradient obfuscation techniques developed in the adversarial examples literature  coupled with randomisation, as suggested in \cite{qiu2020mitigating}. It is well-known that gradient obfuscation may not always prevent an attacker from finding a trigger \cite{athalye2018obfuscated}. Yet, making this process as complicated and computationally-intensive as possible would contribute to increased security.

{\it Pruning to reduce redundancy and the dimension of latent space.} We demonstrate theoretically and confirm in experiments that over-parameterization and high intrinsic dimension of network latent spaces, inherent in many deep learning models, can 
be readily exploited by an adversary if models are freely shared and exchanged without control. In order to deny these opportunities to the attacker, removing susceptible neurons with our procedure in Section \ref{sec:neuron_selection} offers a potential remedy.
More generally, employing network pruning  
\cite{blalock2020state,cheng2017survey,mirkes2020artificial,NEURIPS2020_46a4378f} and enforcing low dimensionality of 
latent spaces as a part of model production pipelines would offer further protection against one neuron attacks.

{\it Network Hashing.} In addition to the strategies above,
which stem from our theoretical analysis of stealth attacks, another defence
mechanism is to use fast network hashing algorithms  executed in parallel with  inference processes.  The hash codes these algorithms produce will enable the owner to detect unwanted changes to the model after it was downloaded from a safe and trusted source.

Our theoretical and practical analysis of the new threats are in no way complete. In our experiments we assumed that pixels in images are real numbers. In some contexts they may take integer values. We did not assess how changes in numerical precision would affect the threat, and did not provide conditions under which redundant neurons exist. Moreover, as we showed in Section \ref{sec:accuracy}, our theoretical bounds could be conservative. The theory presented in this work is fully general in the sense that it does not require any metric in the input space. An interesting question is how one can use the additional structure 
arising when the input space has a metric to find trigger inputs that are closer to their corresponding targets. Finally, vulnerabilities discovered here are intrinsically linked with AI maintenance problems discussed in \cite{GorMakTyu:2018,gorban2018correction, gorban2021high}. Exploring these issues are topics for future research.

Nevertheless, the theory and empirical evidence which we present in this work 
make it clear that 
existing mitigation strategies must be strengthened
in order to 
guard against vulnerability to 
new forms of stealth attack.
Because the ``inevitability'' results that we derived have constructive proofs, our analysis offers promising options for 
the development of effective defences.

\section*{Funding}

This work is supported in part by the UKRI, EPSRC [UKRI Turing AI Fellowship ARaISE EP/V025295/2 and UKRI Trustworthy Autonomous Systems Node in Verifiability EP/V026801/2 to I.Y.T.,  EP/V046527/1 and EP/P020720/1 to D.J.H, and EP/V046527/1 to A.B.].

\bibliographystyle{plain}
\bibliography{adversarial_concentration}


\appendix

\section{Appendix. Proofs of theorems and supplementary results}

\subsection{Proof of Theorem \ref{thm:plain}}

The proof is split into 4 parts. We begin by {\it assuming} that latent representations $\bfx_i=\Phi(\bfu_i)$ of all elements $\bfu_i$ from the set $\mathcal{V}$ belong to the unit ball $\mathbb{B}_n$ centered at the origin (for simplicity of notation 
the unit $n$-ball centered at $0$ is denoted $\mathbb{B}_n$, and the unit $n-1$ sphere centered at $0$ is denoted  $\mathbb{S}_{n-1}$).  The  main thrust of the proof is to establish lower bounds on the probability of the event
\begin{equation}\label{eq:stealth_events}
	\mathcal{E}^\ast: \ \gamma \langle \bfx',\bfx'\rangle \geq \langle \bfx',\bfx_i\rangle \ \mbox{for all} \ \bfx_i=\Phi(\bfu_i): \bfu_i\in\mathcal{V}, \ \bfx_i\in\mathbb{B}_n.
\end{equation}
These bounds are established in {\it Parts 1 and 2} of the proof. Similar events have been shown to play an important role in the problem of AI error correction \cite{gorban2018correction, gorban2021high}---a somewhat dual task to stealth attacks considered here. 

Then we proceed with constructing weights (both, input and output) and biases of the function $g$ so that the modified map $F_a$ delivers a solution of Problem \ref{problem:stealth_plain}. This is shown in {\it Part 3}. Finally, we remove the assumption that $\bfx_i\in\mathbb{B}_n$ and show how the weights and biases need to be adapted so that the resulting adapted map $F_a$ is a solution of Problem \ref{problem:stealth_plain} for $\bfx_i\in\mathbb{B}_n(0,R)$ were $R>0$ is a given number. This is demonstrated in {\it Part 4} of the proof.

\begin{figure}[!h]
	\centering
	\includegraphics[width=0.35\textwidth]{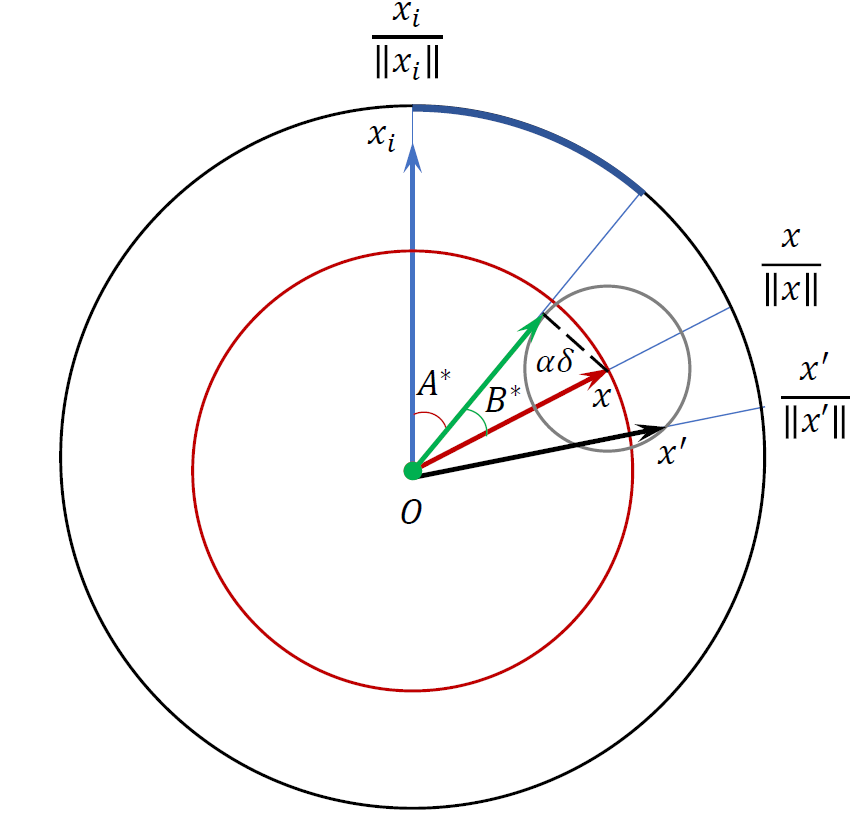} \hspace{5mm}
	\includegraphics[width=0.35\textwidth]{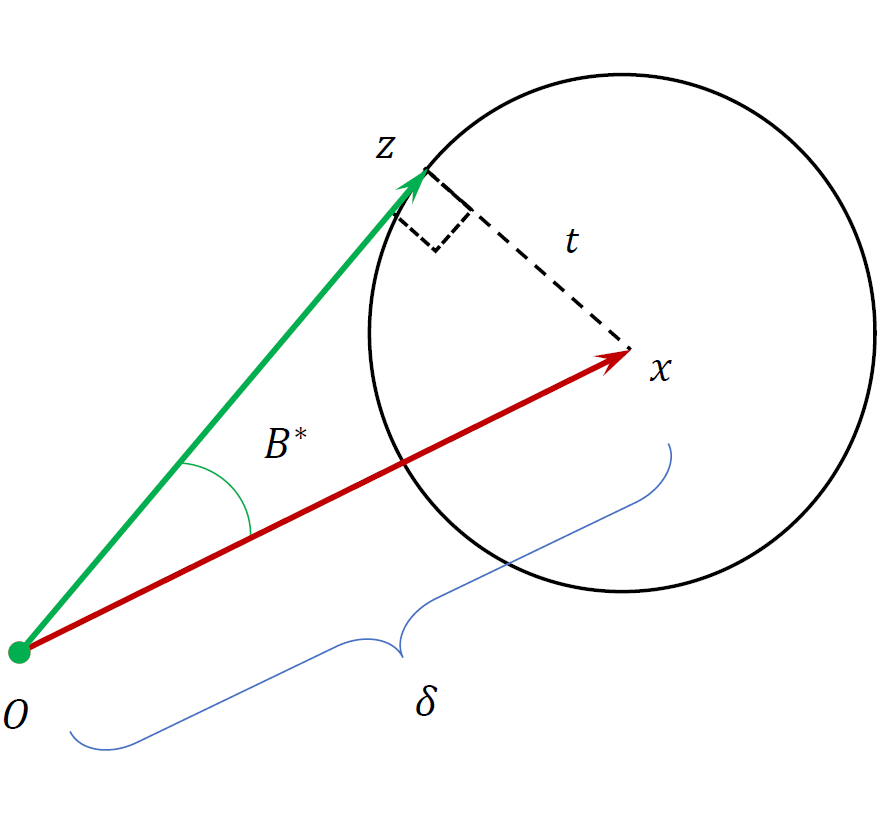}
	\caption{A diagram assisting with the proof of Theorem \ref{thm:plain}. {\it Left panel} illustrates the setup and main ingredients of the argument. {\it Right panel} illustrates the derivation of $B^\ast(x)$. For $t =0$ the expression is trivial. Let $t\neq 0$. Straightforward calculations show that for $\bfz$ corresponding to $\max_{\bfz: \ \|\bfx-\bfz\|=t} |\angle(\bfz,\bfx)|$, the vectors $\bfz$ and $\bfz-\bfx$ must be orthogonal. Hence $\cos \angle(\bfz,\bfx) =\sqrt{\delta^2 - t^2}/\delta$.}\label{fig:proof}
\end{figure}

{\it Part 1. Probability bound 1 on the event $\mathcal{E}^\ast$.} Suppose that $R=1$. Let $\bfx_i$, $i=1,\dots,M$ be an arbitrary element from $\mathcal{V}$,  let $\bfx$ be drawn randomly from an equidistribution in $\mathbb{S}_{n-1}(0,\delta)$, and let $R \bfx'= \Phi(\bfu')$ be a vector within an $\alpha\delta$-distance from $\bfx$:
\[
\|\bfx-\bfx'\|\leq \alpha \delta.
\]
The assumption that the set $\mathbb{S}_{n-1}(0,R\delta)$ is $\delta\alpha R$-input reachable for the map $\Phi$ assures that such $\bfx'$ exists. Moreover, since $\alpha\in[0,1)$, Algorithm $\ref{alg:plain}$ ensures that $\|\bfx'\|\neq 0$ (see (\ref{eq:trigger_condition_1})).

Notice that the constraints $(1+\alpha)\delta\leq  1$ and $\gamma \in [0,1]$ imply that $\gamma \|\bfx'\|\leq \gamma(\|\bfx - \bfx'\| + \|\bfx\|) \leq \gamma(\alpha\delta + \delta) \leq 1$ (i.e. $\arccos(\gamma \|\bfx'\|)$ always exists). Set $
\beta(\gamma,\alpha,\|\bfx'\|):=\arccos(\gamma \|\bfx'\|) + \arccos((1-\alpha^2)^{1/2})$. We claim that if the event
\begin{equation}\label{eq:target_angle_2}
	\mathcal{E}: \ \left(\frac{\bfx}{\|\bfx\|}, \frac{\bfx_i}{\|\bfx_i\|}\right) < \cos \left( \beta(\gamma,\alpha,\|\bfx'\|) \right) \ \mbox{for all} \ \bfx_i\in\mathcal{V}
\end{equation}
occurs then the event $\mathcal{E}^\ast$ (defined in  (\ref{eq:stealth_events})) must occur too. 

To show this, assume that \eqref{eq:target_angle_2} holds and fix $i \in \{1,2,\dotsc,M\}$. By the definition of the dot product in $\mathbb{R}^n$, we obtain
\begin{equation}\label{eq:immediateEventConsequence}
	|\angle(\bfx_i,\bfx)|\geq \beta =  \arccos(\gamma \|\bfx'\|) + \arccos((1-\alpha^2)^{1/2}).
\end{equation}
Consider angles between $\bfx_i,\bfx'$ and $\bfx', \bfx$ as follows: we let
\begin{equation*}
	A^\ast(\bfx,\bfx_i):=\min_{\bfz: \ \|\bfx-\bfz\|\leq \alpha \delta}|\angle(\bfx_i,\bfz)|, 
	\quad B^\ast(\bfx):=\max_{\bfz: \ \|\bfx-\bfz\|\leq \alpha \delta}|\angle(\bfz, \bfx)|.
\end{equation*}
We note that (using the triangle inequality for angular distance)
\begin{equation}\label{eq:AStarBStarBound}
	\begin{split}
		& A^\ast(\bfx,\bfx_i) +  B^\ast(\bfx) \geq |\angle(\bfx_i, \bfx)|
	\end{split}
\end{equation}
(see Figure~\ref{fig:proof}) and  
\begin{equation} \label{eq:BStarValue}B^\ast(\bfx) = \sup_{t \in [0,\alpha\delta]}\, \max_{\bfz: \ \|\bfx-\bfz\|= t}|\angle(\bfz, \bfx)| =  \sup_{t \in [0,\alpha\delta]} \arccos\left(\frac{\sqrt{\delta^2 - t^2}}{\delta}\right) = \arccos(\sqrt{1-\alpha^2}),\end{equation}
where the second equality is illustrated in Figure~\ref{fig:proof} and the final equality follows because $\arccos$ is decreasing.

Combining \eqref{eq:AStarBStarBound}, \eqref{eq:BStarValue} and  \eqref{eq:immediateEventConsequence}  gives
\begin{equation*}
	A^\ast(\bfx,\bfx_i)  \geq |\angle(\bfx_i, \bfx)| -  B^\ast(\bfx) = |\angle(\bfx_i, \bfx)| -\arccos(\sqrt{1-\alpha^2}) \geq \arccos(\gamma \|x'\|)
\end{equation*}
so that (by the fact that cosine is decreasing, the assumption that $\|\bfx_i\| \leq 1$, and noting that $|\angle(\bfx',\bfx_i)| \geq A^\ast(\bfx,\bfx_i)$)
\begin{equation*}
	\langle\bfx', \bfx_i\rangle = \|\bfx'\| \|\bfx_i\|\cos(\angle(\bfx',\bfx_i)) \leq  \|\bfx'\| \cos(A^\ast(\bfx,\bfx_i)) \leq \gamma \langle \bfx',\bfx'\rangle \end{equation*}
completing the proof that if the event \eqref{eq:target_angle_2} occurs then the event \eqref{eq:stealth_events} must occur too.

Consider events
\[
\mathcal{E}_i: \  \langle \frac{\bfx}{\|\bfx\|}, \frac{\bfx_i}{\|\bfx_i\|}\rangle   \geq \cos(\beta(\gamma,\alpha,\|\bfx'\|)).
\]
The probability that $\mathcal{E}_i$ occurs is equal to the area of the spherical cap
\[
C(\bfx_i, \beta(\gamma,\alpha,\|\bfx'\|)) =\left\{\bfz\in\mathbb{S}_{n-1} \left| \ \langle \frac{\bfx_i}{\|\bfx_i\|}, \bfz \rangle \geq \cos(\beta(\gamma,\alpha,\|\bfx'\|)) \right. \right\}
\]
divided by the area of $\mathbb{S}_{n-1}$:
\begin{equation}\label{eq:probability_area}
	P(\mathcal{E}_i) = \frac{A_{n-1}(C(\bfx_i,\beta(\gamma,\alpha,\|\bfx'\|)))}{A_{n-1}(\mathbb{S}_{n-1})}
\end{equation}
(here $A_{n-1}(\mathbb{S}_{n-1})$ stands for the area of the unit $n-1$ sphere $\mathbb{S}_{n-1}$ and $A_{n-1}(C(\bfx_i, \beta(\gamma,\alpha,\|\bfx'\|)))$ denotes the area of the spherical cap $C(\bfx_i, \beta(\gamma,\alpha,\|\bfx'\|))$). It is well-known that 
\[
A_{n-1}(C(\bfx_i,\beta(\gamma,\alpha,\|\bfx'\|)))=A_{n-2}(\mathbb{S}_{n-2}) \int_0^{\beta(\gamma,\alpha,\|\bfx'\|)} \sin^{n-2}(\theta)d\theta, \ \mbox{and} \ A_{n-1}(\mathbb{S}_{n-1})=\frac{2 \pi^{\frac{n}{2}}}{\Gamma(\frac{n}{2})}.
\]
Hence
\[
P\left(\mathcal{E}_i\right) = \frac{A_{n-2}(\mathbb{S}_{n-2})}{A_{n-1}(\mathbb{S}_{n-1})} \int_0^{\beta(\gamma,\alpha,\|\bfx'\|)} \sin^{n-2}(\theta)d\theta.
\]

Using the fact that the inequality
\begin{equation}\label{eq:union_demorgan}
	P(\mbox{not} \ \mathcal{E}_1 \wedge  \cdots  \wedge \mbox{not} \ \mathcal{E}_M) \geq 1 - \sum_{i=1}^M P(\mathcal{E}_i)
\end{equation}
holds true for any events $\mathcal{E}_i$, that $\mathcal{E}=\mbox{not} \ \mathcal{E}_1 \wedge  \cdots  \wedge \mbox{not} \ \mathcal{E}_M$,
and that  $P(\mathcal{E}^\ast) \geq P(\mathcal{E}) $ (as $\mathcal{E}$ implies $\mathcal{E}^\ast$), we can conclude that
\[
P\left(\langle \bfx', \bfx_i \rangle < \gamma \langle \bfx',\bfx' \rangle \ \mbox{for all} \ \bfx_i\in\mathcal{V} \ \right) \geq 1 - M \frac{A_{n-2}(\mathbb{S}_{n-2})}{A_{n-1}(\mathbb{S}_{n-1})} \int_0^{\beta(\gamma,\alpha,\|\bfx'\|)} \sin^{n-2}(\theta)d\theta, 
\]
or, equivalently, 
\begin{equation}\label{eq:proof:theorem:1:bound1}
	\begin{split}
		&P\left(\langle\bfx', \bfx_i \rangle < \gamma \langle \bfx',\bfx' \rangle \ \mbox{for all} \ \bfx_i\in\mathcal{V} \ \right) \geq 1 - M\frac{1}{\pi^{\frac{1}{2}}} \frac{\Gamma\left(\frac{n}{2}\right)}{\Gamma\left(\frac{n-1}{2}\right)} \int_{0}^{\beta(\gamma,\alpha,\|\bfx'\|)} \sin^{n-2}(\theta)d\theta.
	\end{split}
\end{equation}
Finally, noticing that
\[
\beta(\gamma,\alpha,\|\bfx'\|)\leq \arccos(\gamma (1-\alpha)\delta) + \arccos((1-\alpha^2)^{1/2})
\]
we can conclude that Algorithm~\ref{alg:plain} ensures that event $\mathcal{E}^\ast$ in (\ref{eq:stealth_events}) occurs with probability at least (\ref{eq:probability_plain}).

{\it Part 2. Proving the bound \eqref{eq:probability_plain_bound}} 
Since $\cos$ is non-negative and decreasing on $[0,\pi/2]$ and $\arccos( \varphi(\gamma,\delta,\alpha)) \in [0,\pi/2]$ we have $\cos(\theta)/\varphi(\gamma,\delta,\alpha) = \cos(\theta)/\cos(\arccos( \varphi(\gamma,\delta,\alpha))) \geq 1$ for every $\theta \in [0,\arccos( \varphi(\gamma,\delta,\alpha))]$. Hence
\begin{align*}
	\frac{1}{\Gamma\left(\frac{n-1}{2}\right)} \int_{0}^{\arccos{( \varphi(\gamma,\delta,\alpha) )}} \sin^{n-2}(\theta)d\theta &\leq \frac{1}{\Gamma\left(\frac{n-1}{2}\right)} \int_{0}^{\arccos{( \varphi(\gamma,\delta,\alpha) )}} \frac{\cos(\theta)}{\varphi(\gamma,\delta,\alpha)}\sin^{n-2}(\theta)d\theta\\
	&= \frac{1}{\Gamma\left(\frac{n-1}{2}\right)} \frac{1}{\varphi(\gamma,\delta,\alpha)} \left[ \frac{\sin^{n-1}(\arccos(\varphi(\gamma,\delta,\alpha)))}{n-1} -\frac{\sin^{n-1}(0)}{n-1}  \right]
	\\&=   \frac{1}{2\Gamma\left(\frac{n}{2}+\frac{1}{2}\right)} \frac{1}{\varphi(\gamma,\delta,\alpha)} \left(1-\varphi(\gamma,\delta,\alpha)^2\right)^{\frac{n-1}{2}}.
\end{align*}
As a result,
\begin{equation}\label{eq:proof:theorem1:bound:2}
	P(\mathcal{E}_i) \leq  \frac{1}{2 \pi^{\frac{1}{2}}} \frac{\Gamma\left(\frac{n}{2}\right)}{\Gamma\left(\frac{n}{2}+\frac{1}{2}\right)}  \frac{1}{\varphi(\gamma,\delta,\alpha) } \left(1-\varphi(\gamma,\delta,\alpha)^2\right)^{\frac{n-1}{2}},
\end{equation}
from which we obtain \eqref{eq:probability_plain_bound}.

The same estimate can be obtained via an alternative geometrical argument by recalling that \begin{equation}
	\begin{split}
		&A_{n-1}(C(\bfx_i, \arccos(\varphi(\gamma,\delta,\alpha)) )) \frac{1}{n} = V_n(C(\bfx_i, \arccos(\varphi(\gamma,\delta,\alpha)) ) + \\
		& \ \ \ \ \ \ \ \ \ \ \ \ \  \frac{\varphi(\gamma,\delta,\alpha)}{n} V_{n-1}\left(B_{n-1}\left(\frac{\bfx_i}{\|\bfx_i\|} \varphi(\gamma,\delta,\alpha) , (1-\varphi(\gamma,\delta,\alpha)^2)^{1/2} \right)\right).
	\end{split}\nonumber
\end{equation}
and then estimating the volume of the spherical cap $C(\bfx_i, \arccos(\varphi(\gamma,\delta,\alpha))$ by the volume of the corresponding spherical cone containing $C(\bfx_i, \arccos(\varphi(\gamma,\delta,\alpha))$ whose base is the disc centered at $\frac{\bfx_i}{\|\bfx_i\|} \varphi(\gamma,\delta,\alpha)$ with radius $(1-\varphi(\gamma,\delta,\alpha)^2)^{1/2}$, and whose height is $\frac{1}{\varphi(\gamma,\delta,\alpha)} - \varphi(\gamma,\delta,\alpha)$.

{\it Part 3. Construction of the structural adversarial perturbation.} Having established bounds on the probability of event $\mathcal{E}^\ast$ in (\ref{eq:stealth_events}), let us now proceed with determining a map $F_a$ which is solution to Problem \ref{problem:stealth_plain} {\it assuming} that $\bfx_i\in\mathbb{B}_n$.  Suppose that the event $\mathcal{E}^\ast$ holds true. 

By construction, 
since $R=1$, in Algorithm~\ref{alg:plain} we have 
\begin{eqnarray}\label{eq:attack_parameters_ball}
	\bfw &=& \kappa \bfx',  \ \kappa>0, \\
	b & =& \kappa \left(\frac{1+\gamma}{2}\right) \|\bfx'\|^2,
\end{eqnarray}
and 
\[
\mathfrak{A}(\cdot,\bfw,b)=D g\left(\kappa\left( \langle \cdot,\bfx' \rangle - \left(\frac{1+\gamma}{2}\right) \|\bfx'\|^2 \right)\right).
\]
Since the function $g$ is monotone,
\[
|\mathfrak{A}(\bfx_i,\bfw,b)|\leq D g\left(- \kappa \left(\frac{1-\gamma}{2} \|\bfx'\|^2 \right)\right) \ \mbox{for all} \  \bfx_i=\Phi(\bfu_i), \ \bfu_i\in\mathcal{V}, \ \bfx_i\in\mathbb{B}_n.
\]
Writing
\[
z = \frac{1-\gamma}{2} \|\bfx'\|^2
\]
we 
note from (\ref{eq:adversarial_solution_parameters}) that 
the values of $D$ and $\kappa$ are chosen so that 
\begin{equation}\label{eq:attack_parameters_part_2}
	D g (-\kappa z)  \leq \varepsilon \ \mbox{and} \
	D g (\kappa z )  \geq \Delta.
\end{equation}

{\it Part 4. Generalisation to the case when $\bfx_i\in\mathbb{B}_n(0,R)$.}  Consider variables 
\[
\tilde{\bfx}_i=\frac{\bfx_i}{R}.
\]
It is clear that $\tilde{\bfx}_i\in\mathbb{B}_n$. Suppose now that
\begin{equation}\label{eq:event_final}
	\langle \bfx',\tilde{\bfx}_i \rangle \leq \gamma \|\bfx'\|^2 \ \mbox{for all} \ \tilde{\bfx}_i=\frac{\Phi(\bfu_i)}{R}, \ \bfu_i\in\mathcal{V}, \  \bfx_i=\Phi(\bfu_i)\in\mathbb{B}_n(0,R).
\end{equation}
Probability bounds on the above event have already been established in {\it Parts 1 and 2} of the proof.

In this general case, Algorithm~\ref{alg:plain} uses 
\begin{eqnarray*}
	\bfw &=& \kappa \left(\frac{\bfx'}{R}\right),  \ \kappa>0, \\
	b & =& \kappa \left(\frac{1+\gamma}{2}\right) \|\bfx'\|^2,  
\end{eqnarray*}
and 
\[
\mathfrak{A}(\cdot,\bfw,b)=D g\left(\kappa\left( \langle \cdot \frac{1}{R},\bfx'\rangle - \left(\frac{1+\gamma}{2}\right) \|\bfx'\|^2 \right)\right).
\]
The values of $\kappa$ and $D$ are chosen so that (\ref{eq:attack_parameters_part_2}) in {\it Part 3} holds,
and hence   
\[
|\mathfrak{A}(\Phi(\bfu_i),\bfw,b)| \leq \varepsilon  \ \mbox{for all} \ \bfu_i\in\mathcal{V}
\]
and
\[
\mathfrak{A}(\bfx' R ,\bfw,b)=\mathfrak{A}(\Phi(\bfu'),\bfw,b) \geq \Delta. 
\]
This, together with bounds (\ref{eq:proof:theorem:1:bound1}), (\ref{eq:proof:theorem1:bound:2}) in {\it Parts 1 and 2} on the probability of event (\ref{eq:event_final}) concludes the proof.
$\square$.

\subsection{Proof of Theorem~\ref{thm:incremental}}

Consider an AI system (\ref{eq:classification_map_composition}) which satisfies Assumption \ref{assume:data_ball:targeted}. In addition to this system, consider a ``virtual'' one whose maps $F$ and $\Phi$ are replaced with:
\begin{equation}\label{eq:virtual}
	\tilde{\Phi}(\bfu)=\Phi(\bfu)-\Phi(\bfu^\ast), \  \tilde{F}(\bfx)=F(\bfx+\Phi(\bfu^\ast)).
\end{equation}
According to the definition of $\tilde{F}\circ \tilde{\Phi}$, domains of the definition of $\tilde{F}\circ \tilde{\Phi}$ and ${F}\circ {\Phi}$ coincide, and
\begin{equation}\label{eq:virtual_AI_equivalence}
	\tilde{F}\circ \tilde{\Phi} (\bfu)= F\circ \Phi (\bfu) \ \mbox{for all} \ \bfu\in\mathcal{U}.
\end{equation}
For this virtual system, $\tilde{F}\circ \tilde{\Phi}$, Assumption \ref{assume:data_ball} is satisfied. Moreover, if an input $\bfu'$
\[
\bfx'=\frac{\Phi(\bfu')}{R}-\frac{\Phi(\bfu^\ast)}{R}
\]
satisfies condition (\ref{eq:trigger_condition_1}), 
then these conditions are satisfied for $\bfx'=\tilde{\Phi}(\bfu')/R$ (the converse holds true too), and $\|\bfx'\|\neq 0$. 

According to Theorem \ref{thm:plain}, if we define the stealth attack parameters as in  (\ref{eq:adversarial_solution}), (\ref{eq:adversarial_solution_parameters}) by:
\begin{equation}
	\begin{split}
		& \mathfrak{A}\left(\cdot, \kappa \frac{\bfx'}{R}, b\right)=D g\left(\langle \cdot,\kappa \frac{\bfx'}{R} \rangle - b\right), \\
		& b=\kappa \left(\frac{1+\gamma}{2}\right) \|\bfx'\|^2
	\end{split}\nonumber
\end{equation}
then the above is a solution of Problem \ref{problem:stealth_plain} for the virtual AI system $\tilde{F}\circ \tilde{\Phi}$ with probability bounded from below as in (\ref{eq:probability_plain}), (\ref{eq:probability_plain_bound}). 

Notice that 
\begin{equation}
	\begin{split}
		&\mathfrak{A}\left(\Phi(\bfu)-\Phi(\bfu^\ast), \kappa \frac{\bfx'}{R}, b \right)=\\
		& \ \ \ \ \ \ \ \ \ \ D g\left(\langle \Phi(\bfu),\kappa \frac{\bfx'}{R} \rangle - \left(\kappa \left(\frac{1+\gamma}{2}\right) \|\bfx'\|^2+\langle \Phi(\bfu^\ast),\kappa \frac{\bfx'}{R}\rangle \right)\right),
	\end{split}\nonumber
\end{equation}
which coincides with the attack used in  Algorithm~\ref{alg:incremental}.
Hence, taking  \eqref{eq:virtual_AI_equivalence} into account, we can conclude that Algorithm \ref{alg:incremental} returns a solution to Problem  \ref{problem:stealth_plain} for the original AI system ${F}\circ {\Phi}$ with probabilities of success satisfying (\ref{eq:probability_plain}) and (\ref{eq:probability_plain_bound}). $\square$

\subsection{Proof of Theorem \ref{thm:concentration_collapse}}
In a similar way to the proof of Theorem \ref{thm:incremental}, we begin with considering a virtual system with maps $\tilde\Phi$ and $\tilde F$ defined as in (\ref{eq:virtual}). We observe that Assumption \ref{assume:concentration} implies that Assumption \ref{assume:data_ball:targeted} holds true. As we have seen before, domains of the definition of $\tilde{F}\circ\tilde{\Phi}$ and $F\circ\Phi$ coincide, and Assumption \ref{assume:data_ball:targeted} implies that Assumption \ref{assume:data_ball} holds true for the virtual system. 

As in the proof of Theorem \ref{thm:incremental}, we will consider the vector $\bfx'$
\[
\bfx'=\frac{\Phi(\bfu')}{R}-\frac{\Phi(\bfu^\ast)}{R}=\frac{\tilde{\Phi}(\bfu')}{R}
\]
satisfying condition (\ref{eq:trigger_condition_1}), where $\bfx$ is drawn from an equidistribution on the sphere $\mathbb{S}_{n_p-1}(0,\delta)$. We set $\tilde{\bfc}=\bfc-\Phi(\bfu^\ast),$ and let $\tilde \bfc = \tilde \bfc^1 + \tilde \bfc^2$ where $\tilde \bfc^1 \in \mbox{span} \{h_1,\dots,h_{n_p}\}$ and $\tilde \bfc^2  \perp \mbox{span} \{h_1,\dots,h_{n_p}\}$. 
Our argument proceeds by executing four steps: 
\begin{enumerate}
	\item We begin by showing that $\theta^* \in [0,\pi]$ and $\sqrt{1-(MC)^{-2/n}} \leq \rho(\theta) \leq 1$ for $\theta \in [\theta^*,\pi]$.
	\item We introduce the random variable $\Theta$ defined by
	$
	\Theta= \angle(\bfx,\tilde{\bfc}^1/R)
	$
	and the events
	\begin{equation}\label{eq:thm3:target_event}
		\mbox{for } i=1,\dots,M,\, \mathcal{E}_i^\ast \text{ is the following event}: \left \langle \frac{\bfx'}{\|\bfx'\|}, \bfx_i \right\rangle \leq \gamma \|\bfx'\|, \ \mbox{with } \ \bfx_i=\tilde{\Phi}(\bfu_i)/R.  
	\end{equation}
	and	
	\[
	\mbox{for } i=1,\dots,M,\, \mathcal{E}_{i}(\Theta) \mbox{ is the following event} : \ \frac{\tilde{\Phi}(\bfu_i)}{R}\in \Omega(x', \rho(\Theta)), \ i=1,\dots M,
	\]
	with, for $\theta \in [\theta^*, \pi]$, \[\Omega(z, \rho(\theta)):=\{\bfx\in \mathbb{B}_n(\tilde{\bfc}/R,1/2) \ | \ \langle h(\bfz),\bfx-\tilde{\bfc}/R\rangle \leq \rho(\theta)/2  \}, \quad h(z)  := z/\|z\|_2.\]
	In the second step, we condition on $\Theta = \theta$ with $\theta \geq \theta^*$ and show that if
	$\mathcal{E}_1(\Theta)\wedge \cdots \wedge \mathcal{E}_M(\Theta)|  \Theta=\theta$ occurs then so too does the event $\mathcal{E}_1^\ast\wedge\cdots \wedge \mathcal{E}_M^\ast | \Theta=\theta$.
	\item The third step consists of producing a lower bound on the  $P(\mathcal{E}_1(\Theta)\wedge \cdots \wedge \mathcal{E}_M(\Theta)|  \Theta=\theta)$ for $\theta \in [\theta^*,\pi]$. This is done by relating the event  $\mathcal{E}_{i}|\Theta=\theta$ to Assumption \ref{assume:concentration}.
	
	\item Finally, we use the previous steps to prove that the event $\mathcal{E}_1^\ast\wedge\cdots \wedge \mathcal{E}_M^\ast$ happens with at least the desired probability. Then, under the assumption that $\mathcal{E}_1^\ast\wedge\cdots \wedge \mathcal{E}_M^\ast$ occurs, a simple argument similar to the one presented in the final two paragraphs of Theorem 2 allows us to conclude the argument.

	 
\end{enumerate}
\textit{Step 1: Proving that $\theta^* \in [0,\pi]$ and $\sqrt{1-(MC)^{-2/n}} \leq \rho(\theta) \leq 1$ for $\theta \in [\theta^*,\pi]$.}

We start with the claim that $\theta^* \in [0,\pi]$. Since $\alpha \in [0,1]$, we must have $\arccos(\sqrt{1-\alpha^2}) \in [0,\pi/2]$. Moreover, by the assumption $2\delta\gamma(1-\alpha) \geq \sqrt{1-(MC)^{-2/n}}$, we must also have  \[\arccos\left(\min\left(1,2\delta\gamma(1-\alpha) - \sqrt{1-(MC)^{-2/n}}\right)\right) \in [0,\pi/2].\] The result follows.

Next, we show that $\sqrt{1-(MC)^{-2/n}} \leq \rho(\theta) \leq 1$ for $\theta \in [\theta^*,\pi]$. The upper bound is obvious from the definition of $\rho(\theta)$. For the lower bound, the function $\theta: \to -\max\{0,\cos (\theta - \arccos(1-\alpha^2)^{1/2})$ is an increasing function of $\theta$ for $\theta \in [\arccos(\sqrt{1-\alpha^2}),\pi]$ and hence $\rho(\theta)$ is increasing on $[\theta^*,\pi]$. Therefore it suffices to prove that $\rho(\theta^*) \geq \sqrt{1-(MC)^{-2/n}}$. To do this, note that
\begin{align*} \rho(\theta^*) &= \min\left\lbrace 1,2\delta \gamma (1-\alpha) - \max\left\{0,\min\left(1,2\delta\gamma(1-\alpha) - \sqrt{1-(MC)^{-2/n}}\right)\right\} \right\rbrace \\
	&= \min\left\lbrace 1,2\delta \gamma (1-\alpha) - \min\left(1,2\delta\gamma(1-\alpha) - \sqrt{1-(MC)^{-2/n}}\right) \right\rbrace \\
	&= \min\left\lbrace 1, \max\left(2\delta \gamma (1-\alpha)-1,2\delta \gamma (1-\alpha)-2\delta\gamma(1-\alpha) + \sqrt{1-(MC)^{-2/n}}\right) \right\rbrace\\
	&= \min\left\lbrace 1, \max\left(2\delta \gamma (1-\alpha)-1,\sqrt{1-(MC)^{-2/n}}\right) \right\rbrace 
	\\&\geq \min\left\lbrace 1, \sqrt{1-(MC)^{-2/n}} \right\rbrace = \sqrt{1-(MC)^{-2/n}},
\end{align*}
where we have used the assumption $2\delta\gamma(1-\alpha) \geq \sqrt{1-(MC)^{-2/n}}$ in the second equality, we have used that $\lambda_1 - \min(\lambda_2,\lambda_3) = \max(\lambda_1 - \lambda_2,\lambda_1 - \lambda_3)$ for real numbers $\lambda_1,\lambda_2,\lambda_3$ in the third equality, and in the inequality we use the fact that both $\min$ and $\max$ are increasing functions of their arguments.

\textit{Step 2: Showing that if the event $\mathcal{E}_1(\Theta)\wedge \cdots \wedge \mathcal{E}_M(\Theta)|  \Theta=\theta$ occurs then the event $\mathcal{E}_1^\ast\wedge\cdots \wedge \mathcal{E}_M^\ast | \Theta=\theta$ also occurs for $\theta \in [\theta^*,\pi]$.}

First, suppose that the event  $\mathcal{E}_1(\Theta)\wedge \cdots \wedge \mathcal{E}_M(\Theta)|  \Theta=\theta$ occurs. Then for $i \in \{1,2,\dotsc,M\}$, $\frac{\tilde{\Phi}(\bfu_i)}{R}\in \Omega(x', \rho(\theta))$ so that 
we must have $\langle h(\bfx'),\tilde{\Phi}(\bfu_i)/R-\tilde{\bfc}/R \rangle \leq \rho(\theta)/2$. Hence
\begin{equation}\label{eq:Theorem3Step2Int1}
	\langle h(\bfx'),\tilde{\Phi}(\bfu_i)/R \rangle = \langle h(\bfx'),\tilde{\bfc}/R \rangle + \langle h(\bfx'),\tilde{\Phi}(\bfu_i)/R-\tilde{\bfc}/R \rangle  \leq \langle h(\bfx'),\tilde{\bfc}/R \rangle + \rho(\theta)/2.
\end{equation}
Note that $x' \in \mbox{span} \{h_1,\dots,h_{n_p}\}$ and $\tilde \bfc^2  \perp \mbox{span} \{h_1,\dots,h_{n_p}\}$ so that $\langle h(x'), \tilde c/R \rangle = \langle h(x'),\tilde c^1/R\rangle$. The same reasoning tells us that $\|\tilde \bfc^1/R\| \leq \|\tilde{\bfc}/R\|\leq 1/2$ (where the final inequality uses that $\tilde \bfc = c - \Phi(u^*)$ and Assumption \ref{assume:concentration}, \eqref{eq:support}) and hence
\begin{equation}\label{eq:Theorem3Step2Int2}
	\langle h(\bfx'),\tilde{\bfc}/R\rangle = \|h(\bfx')\| \|\tilde{\bfc}^1/R\| \cos(\angle(h(\bfx'),\tilde{\bfc}^1/R)) \leq  \max\{0,\cos(\angle(h(\bfx'),\tilde{\bfc}^1/R))\}/2.
\end{equation}
Combining \eqref{eq:Theorem3Step2Int1} and \eqref{eq:Theorem3Step2Int2} gives us
\begin{equation}\label{eq:thm3:intermediate_step}
	\langle h(\bfx'),\tilde{\Phi}(\bfu_i)/R\rangle \leq \frac{\max\{0,\cos(\angle (h(\bfx'),\tilde{\bfc}^1/R))\} + \rho(\theta)}{2} .
\end{equation}

Next, recall that (using an argument similar to that in Figure \ref{fig:proof} and equations \eqref{eq:AStarBStarBound},\eqref{eq:BStarValue})
$
\cos(\angle (h(\bfx'),\tilde{\bfc}^1/R)) \leq \cos(\Theta - \arccos((1-\alpha^2)^{1/2}))).
$
Noticing that $\|\bfx'\|\neq 0$ (since $\alpha\in[0,1)$) and using (\ref{eq:thm3:intermediate_step}) we see that
\begin{align*}
	\langle \frac{\tilde{\Phi}(\bfu_i)}{R}, \frac{\bfx'}{\|\bfx'\|} \rangle &\leq \frac{1}{2}\rho(\Theta) + \frac{1}{2} \max\{0,\cos(\angle (h(\bfx'),\tilde{\bfc}^1/R))\} \\&\leq \frac{1}{2}\rho(\Theta) + \frac{1}{2} \max\{0,\cos(\Theta - \arccos((1-\alpha^2)^{1/2}))) \} 
	\\&\leq  \delta \gamma (1-\alpha) - \frac{1}{2} \max\{0,\cos (\Theta - \arccos(1-\alpha^2)^{1/2})\}  \\ 
	& + \frac{1}{2} \max\{0,\cos(\Theta - \arccos((1-\alpha^2)^{1/2}))) \}  \\&= \delta \gamma (1-\alpha) \leq \gamma \|\bfx'\|, 
\end{align*}
where the final inequality follows because $\|\bfx' - \bfx\| \leq \alpha \delta$ and $\|\bfx\| = \delta$. Thus $\mathcal{E}_1^\ast\wedge\cdots \wedge \mathcal{E}_M^\ast | \Theta=\theta$ also occurs and so we conclude that
\begin{equation}\label{EimpliesE^*}
	\mathcal{E}_1(\Theta)\wedge \cdots \wedge \mathcal{E}_M(\Theta)|  \Theta=\theta \implies \mathcal{E}_1^\ast\wedge\cdots \wedge \mathcal{E}_M^\ast | \Theta=\theta \text{ whenever } \theta \in [\theta^*,\pi]
\end{equation}
\textit{Step 3: Bounding
$P(\mathcal{E}_1(\Theta)\wedge \cdots \wedge \mathcal{E}_M(\Theta)|\Theta=\theta)$ from below.} 

Assumption \ref{assume:concentration}, equation \eqref{eq:support} implies that all $\Phi(\bfu_i)$ belong to the ball of radius $R/2$ centred at $\bfc$. Hence according to the equivalence 
\[
\|\Phi(\bfu_i)-\bfc\|\leq \frac{R}{2} \ \Leftrightarrow \left\|\frac{\tilde\Phi(\bfu_i)}{R} - \frac{\tilde\bfc}{R}\right\| = \left\|\frac{\Phi(\bfu_i)-\Phi(\bfu^\ast)}{R} - \frac{\bfc - \Phi(\bfu^\ast)}{R}\right\| \leq \frac{1}{2},
\]
we conclude that all $\tilde{\Phi}(\bfu_i)/R$ belong to the  ball $\mathbb{B}_n(\tilde\bfc/R,1/2)$ of radius $1/2$ centred at  $\tilde{\bfc}/R$ where we recall that $\tilde \bfc=\bfc-\Phi(\bfu^\ast)$.

Thus for any $i=1,\dots,M$ the probability $P(\mbox{not }\mathcal{E}_{i}| \Theta=\theta)$, $\theta\in[\theta^*,\pi]$ is the probability of the random point $\tilde{\Phi}(\bfu_i)/R$ landing in the spherical cap
\[
C^\ast=\left\{\bfy \in \mathbb{B}_n(\tilde\bfc/R,1/2) \left| \ \langle \frac{\bfx'}{\|\bfx'\|}, \bfy -\tilde{\bfc}/R \rangle \geq \rho(\theta)/2  \right. \right\}.
\]
Any point $\bfy$ in the set $C^\ast$ satisfies:
\[
\begin{split}
&\left\|\frac{\tilde{\bfc}}{R}+ \frac{\bfx'}{2\|\bfx'\|}\rho(\theta) - \bfy\right\|^2 \leq 
\|\bfy-\frac{\tilde{\bfc}}{R}\|^2 - \rho(\theta) \left\langle \frac{\bfx'}{\|\bfx'\|}, \bfy -\frac{\tilde{\bfc}}{R} \right\rangle +   \frac{\rho(\theta)^2}{4}\leq \\
&  \|\bfy-\frac{\tilde{\bfc}}{R}\|^2 - \frac{\rho(\theta)^2}{4} \leq \frac{1- \rho(\theta)^2}{4}.
\end{split}
\]

In particular, if $\tilde \Phi(\bfu_i)/R \in C^{\ast}$ and we define $\xi = \bfc + R h(\bfx') \rho(\theta)/2$ and $r = R(1 - \rho(\theta)^2)^{1/2}/2$ (where $r$ is real since $\rho(\theta) \in [0,1/2]$) then
\[
\begin{split}
\|\xi -\Phi(u_i)\|^2 &=\|\tilde \bfc + Rh(\bfx') \rho(\theta)/2  -\tilde \Phi(u_i)\|^2 = R^{2}\left\|\frac{\tilde{\bfc}}{R}+ \frac{\bfx'}{2\|\bfx'\|}\rho(\theta) - \frac{\tilde \Phi(\bfu_i)}{R} \right\|^2 \\
&  \leq  \frac{R^2(1 - \rho(\theta)^2)}{4} = r^2,
\end{split}
\]
so that \[P(\mbox{not }\mathcal{E}_{i}| \Theta=\theta) = P(\Phi(\tilde \bfu_i)/R \in C^{\ast} | \Theta = \theta) \leq P(\Phi(\bfu_i) \in \mathbb{B}_n(\xi,r)|\Theta = \theta). \]

Next, note that $\|\xi - c\| = R |\rho(\theta)|/2 \leq R/2$ since $\rho(\theta) \in [0,1]$ for $\theta \in [\theta^*,\pi]$. Also, $r \leq R/2$. Moreover, for any given value of $\theta$ the values $\xi$ and $r$ are deterministic and $\Theta$ is independent of $\Phi(\bfu_i)$ since $x$ was chosen in the algorithm without knowledge of $\bfu_i$. Therefore we can apply Assumption \ref{assume:concentration}, equation \eqref{eq:exponent_concentration} to obtain
\[
P(\mbox{not }\mathcal{E}_{i}|\Theta=\theta) \leq C \left(2\left(\frac{1 - \rho(\theta)^2}{4} \right)^\frac{1}{2}\right)^n = C \left(1-\rho(\theta)^2 \right)^\frac{n}{2}   \ \mbox{for all} \ i=1,\dots,M.
\]
Therefore 
\begin{equation}\label{eq:E_iboundfrombelow}
P(\mathcal{E}_1\wedge \cdots \wedge \mathcal{E}_{M}|\Theta=\theta) \geq 1 - \sum_{i=1}^{M} P(\mbox{not }\mathcal{E}_i|\Theta=\theta)\geq 1 - M C \left(1-\rho(\theta)^2 \right)^\frac{n}{2}
\end{equation}
\textit{Step 4: Concluding the argument}

Given that $\bfx = \sum_{i=1}^{n_p} \bfv_i h_i$ with $\bfv$ drawn from the equidistribution in $\mathbb{S}_{n_p-1}(0,\delta)$, the variable $\Theta$ admits the probability density $f_{\Theta}$ with
\begin{equation} \label{eq:ThetaDensity}
f_{\Theta}(\theta) = \frac{1}{\pi^{\frac{1}{2}}} \frac{\Gamma\left(\frac{n_p}{2}\right)}{\Gamma\left(\frac{n_p-1}{2}\right)} \sin^{n_p-2}(\theta),  \      \theta \in [0,\pi].
\end{equation}
Therefore
\[
\begin{split}
	P(\mathcal{E}_1^\ast\wedge \cdots \wedge \mathcal{E}_M^\ast)=&\int_0^{\pi}  P(\mathcal{E}_1^\ast\wedge \cdots \wedge \mathcal{E}_M^\ast|\Theta=\theta) f_{\Theta}(\theta) d\theta\geq \int_{\theta^*}^{\pi}  P(\mathcal{E}_1^\ast\wedge \cdots \wedge \mathcal{E}_M^\ast|\Theta=\theta) f_{\Theta}(\theta) d\theta\\
	\geq & \int_{\theta^*}^{\pi}  P(\mathcal{E}_1(\Theta)\wedge \cdots \wedge \mathcal{E}_M(\Theta)|\Theta=\theta) f_{\Theta}(\theta) d\theta.
\end{split}
\]
where the final inequality follows from the \eqref{EimpliesE^*} proven in step 2.

In step 1 we showed that $\rho(\theta)\geq \sqrt{1-(MC)^{-2/n}}$ for $\theta \in [\theta^*,\pi]$. Rearranging yields $1  - MC(1-\rho(\theta)^2)^{n/2} \geq 0$. Combining this with the inequality \eqref{eq:E_iboundfrombelow} we obtain 
\begin{align*}
	P(\mathcal{E}_1^\ast\wedge \cdots \wedge \mathcal{E}_{M}^\ast)&\geq 
	\int_{\theta^\ast}^{\pi} (1-MC \left(1-\rho(\theta)^2 \right)^\frac{n}{2} )f_{\Theta}(\theta)d\theta 
	\\& = \int_{\theta^*}^{\pi} f_{\Theta}(\theta)d\theta  - M C \int_{\theta^\ast}^{\pi}  \left(1-\rho(\theta)^2 \right)^\frac{n}{2}f_{\Theta}(\theta)d\theta \\& = 1 - M C \int_{\theta^\ast}^{\pi}  \left(1-\rho(\theta)^2 \right)^\frac{n}{2}f_{\Theta}(\theta)d\theta -  \int_0^{\theta^\ast} f_{\Theta}(\theta)d\theta.  
\end{align*}
Substituting \eqref{eq:ThetaDensity} gives
\[
\begin{split}
	P(\mathcal{E}_1^\ast\wedge \cdots \wedge \mathcal{E}_{M}^\ast)\geq& 1 - M C \frac{1}{\pi^{\frac{1}{2}}} \frac{\Gamma\left(\frac{n_p}{2}\right)}{\Gamma\left(\frac{n_p-1}{2}\right)} \int_{\theta^\ast}^{\pi}  \left(1-\rho(\theta)^2 \right)^\frac{n}{2} \sin^{n_p-2}(\theta)d\theta \\
	& -  \frac{1}{\pi^{\frac{1}{2}}} \frac{\Gamma\left(\frac{n_p}{2}\right)}{\Gamma\left(\frac{n_p-1}{2}\right)} \int_0^{\theta^\ast} \sin^{n_p-2}(\theta)d\theta.   
\end{split}
\]
Finally, assuming that the event $\mathcal{E}_1^\ast\wedge \cdots \wedge \mathcal{E}_{M}^\ast$ occurs, the final two paragraphs of the argument presented in the proof of Theorem \ref{thm:incremental} (noting that Assumption \ref{assume:concentration} implies \ref{assume:data_ball:targeted}) may then be used to confirm that the values of $\bfw$ and $b$ specified in Algorithm \ref{alg:constrained} produce an $\varepsilon$-$\Delta$ stealth attack. The conclusion of Theorem \ref{thm:concentration_collapse} follows.
$\square$

\subsection{Finding triggers $\bfu'$ in Algorithms \ref{alg:plain} and \ref{alg:incremental}}\label{sec:choosing_triggers}

A relevant optimization problem for Algorithm~\ref{alg:plain} is formulated in (\ref{eq:finding_trigger_1}). Similarly to (\ref{eq:finding_trigger_1}), one can pose a constrained optimization problem for finding a $\bfu'$ in step 3 of Algorithm \ref{alg:incremental}. This problem can be formulated as follows:
\begin{equation}\label{eq:finding_trigger_2}
	\begin{split}
		\bfu'=& \arg \min_{\bfu\in\mathcal{U}} \left\| \frac{\Phi(\bfu)}{R} - \frac{\Phi(\bfu^\ast)}{R}- \bfx \right\| \\
		& \mbox{s.t.} \\
		& \gamma \left\| \frac{\Phi(\bfu)}{R} -\frac{\Phi(\bfu^\ast)}{R}\right\| \leq 1, \quad   \left\|\frac{\Phi(\bfu)}{R} - \frac{\Phi(\bfu^\ast)}{R} - \bfx \right\| < \delta.
	\end{split}
\end{equation}

Note that the vector $\bfx$ must be chosen randomly on $\mathcal{S}_{n-1}(0,\delta)$, $\delta\in(0,1]$. One way to achieve this is to generate a sample $\bfz$ from an $n$-dimensional normal distribution $\mathcal{N}(0,I_n)$ and then set $\bfx=\delta \bfz/\|\bfz\|$.

A practical approach to determine $\bfu'$ is to employ a gradient-based search
\[
\bfu'_{k+1}=\mbox{Proj}_{\mathcal{U}}\left[\bfu'_{k} - h_k \frac{\partial}{\partial \bfu}\mathcal{L}(\bfu,\bfu^\ast)\right], \quad \bfu'_0=\bfu^\ast,
\]
where $\mbox{Proj}_{\mathcal{U}}$ is a projection operator. In our experiments, $\mbox{Proj}_{\mathcal{U}}(\bfu)$ returned $\bfu=(u_1,\dots,u_n)$ if $u_i\in[0,255]$. If, however, the $i$-th component of $\bfu$ is out of range ($u_i<0$ or $u_i>255$) then the operator returned a vector with $0$ or $255$ in that corresponding component.

The loss function $\mathcal{L}(\bfu,\bfu^\ast)$ is defined as
\[
\mathcal{L}(\bfu,\bfu^\ast)=\left\| \frac{\Phi(\bfu)}{R} - \frac{\Phi(\bfu^\ast)}{R}- \bfx \right\|^2 + \lambda_1 \mathcal{G}_1(\bfu) +  \lambda_2\mathcal{G}_2(\bfu), \quad \lambda_1, \ \lambda_2\geq 0,
\]
and $\mathcal{G}_1$, $\mathcal{G}_2$ are relevant penalty functions:
\[
\mathcal{G}_1(\bfu)= \left\{ \begin{array}{ll}
	\left(\gamma \left\| \frac{\Phi(\bfu)}{R} -\frac{\Phi(\bfu^\ast)}{R}\right\| - 1 \right)^{p_1}, & \gamma \left\| \frac{\Phi(\bfu)}{R} -\frac{\Phi(\bfu^\ast)}{R}\right\| \geq 1\\
	0, & \mbox{otherwise}
\end{array}\right.
\]
\[
\mathcal{G}_2(\bfu)= \left\{ \begin{array}{ll}
	\left(\left\|\frac{\Phi(\bfu)}{R} - \frac{\Phi(\bfu^\ast)}{R} - \bfx \right\| -\delta\right)^{p_2}, &  \left\|\frac{\Phi(\bfu)}{R} - \frac{\Phi(\bfu^\ast)}{R} - \bfx \right\| \geq \delta\\
	0, & \mbox{otherwise}
\end{array}\right.
\]
with parameters $p_1,p_2>0$, and with $h_k>0$ a sequence of parameters ensuring convergence of the procedure. 

For Algorithm \ref{alg:plain}, terms $\Phi(\bfu^\ast)/R$ should be replaced with $0$, and $\bfu_0$ can be set to an arbitrary element of $\mathcal{U}$.

\begin{remark}[Trigger search subspace]\label{rem:perturbation_dimension}\normalfont Sometimes we wish to 
	transform the original feature map $\Phi$ into $\tilde\Phi=T \Phi$ (see Remark \ref{rem:input_reachability}) in order to comply with the input-reachability assumption. This translates into sampling  $\bfx$ in step 2 of Algorithm \ref{alg:incremental}  from  $S_{m-1}(0,\delta)\subset \Real^n$, $2\leq m<n$ instead of $S_{n-1}(0,\delta)$.  On the one hand this may have a potentially negative affect on the probability of success as $n$ would need to be replaced with $2\leq m<n$ in (\ref{eq:probability_plain}) and (\ref{eq:probability_plain_bound}). On the other hand this extra ``pre-processing'' may offer computational benefits. Indeed, if $\Phi(\bfu^\ast)$ is an output of a ReLU layer then it is likely that some of its components are exactly zero. This would severely constrain gradient-based routines to determine $\bfu'$ starting from $\bfu^\ast$ if $\bfx$ is sampled from $S_{n-1}(0,\delta)$, as the corresponding gradients will always be equal to zero. In addition, one can constrain the triggers' search space to smaller dimensional subspaces, as specified in Algorithm \ref{alg:constrained}. Not only may this approach relax input-reachability restrictions and help with computing triggers but it may also be used to mitigate various feasibility issues around accuracy of the attacks (see Theorem \ref{thm:concentration_collapse} and Tables \ref{tab:collapse_illustration}, \ref{tab:collapse_illustration_reduced}).
	
	In our code \cite{ExampleCode} illustrating the application of Algorithms \ref{alg:plain}, \ref{alg:incremental}, \ref{alg:constrained} we follow the above logic and search for input triggers in subspaces of the original feature space corresponding to non-zero attributes of the feature vector $\Phi(\bfu^\ast)$.
\end{remark}

\subsection{Accuracy of implementation of stealth attacks achieved in experiments and limits of theoretical bounds in 
	Theorems~\ref{thm:plain} and \ref{thm:incremental}}\label{sec:accuracy}

In all $20$ attack experiments we were able to create a ReLU neuron which was completely silent on the set $\mathcal{V}$ and, at the same time, was producing the desired responses when a trigger was presented as an input to the network. Empirical values of the accuracy of implementation of  these attacks, $\alpha$ and the dimension $n$ of the random perturbation are shown in Table \ref{tab:accuracy}.

\begin{table}
	\centering
	\small
	\begin{tabular}{|c|c|c|c|c|c|c|c|c|c|c|}
		\hline
		$\#$ & 1 & 2 & 3 & 4 & 5 & 6 & 7 & 8 & 9 & 10 \\
		\hline\hline
		$\alpha$    & 0.179 & 0.200 & 0.358 & 0.247 & 0.313 &  0.406 &  0.311 & 0.277 & 0.362 & 0.244 \\
		$n$    & 112 & 125 & 122 & 116 & 125 &  122 &  115 & 121 & 128 & 111 \\
		\hline
	\end{tabular} 
	
	\vspace{5mm}
	
	\begin{tabular}{|c|c|c|c|c|c|c|c|c|c|c|}
		\hline
		$\#$ & 11 & 12 & 13 & 14 & 15 & 16 & 17 & 18 & 19 & 20 \\
		\hline\hline
		$\alpha$    & 0.417 & 0.271 & 0.271 & 0.227 & 0.246 &  0.287 &  0.401 & 0.327 & 0.436 & 0.285 \\
		$n$    & 122 & 124 & 124 & 119 & 110 &  117 & 120 & 109 & 110 & 117\\
		\hline
	\end{tabular} 
	\vspace{5mm}
	
	\caption{Accuracy $\alpha$ of finding triggers expressed as $\alpha=\delta^{-1} \|\Phi(\bfu')/R-\bfx\|$, $\delta=1/3$.}\label{tab:accuracy}
\end{table}

\begin{table}
	\centering
	\small
	\begin{tabular}{|c|c|c|c|c|c|c|c|c|c|c|}
		\hline
		$\#$ & 1 & 2 & 3 & 4 & 5 & 6 & 7 & 8 & 9 & 10 \\
		\hline\hline
		$\alpha$    & 0.028 & 0.004 & 0.092 & 4.3$\times 10^{-4}$ & 0.061 &  0.026 &  0.040 & 6.6$\times10^{-5}$ & 2.7$\times 10^{-4}$ & 0.117 \\
		$n$    & 39 & 38 & 35 & 32 & 32 &  35 & 38 & 33 & 37 & 33 \\
		\hline
	\end{tabular} 
	
	\vspace{5mm}
	
	\begin{tabular}{|c|c|c|c|c|c|c|c|c|c|c|}
		\hline
		$\#$ & 11 & 12 & 13 & $\hspace{3mm}$ 14 $\hspace{3mm}$ & 15 & 16 & 17 & 18 & 19 & 20 \\
		\hline\hline
		$\alpha$    & 0.021 & 0.003 & 0.017 & 0.109 & 0.003 &  0.001 &  2.5$\times 10^{-4}$ & 9.2$\times 10^{-5}$ & 0.022 & 0.011 \\
		$n$    & 35 & 36 & 32 & 38 & 35 &  34 & 34 & 32 & 35 & 33\\
		\hline
	\end{tabular} 
	\vspace{5mm}
	
	\caption{Accuracy $\alpha$ of finding triggers expressed as $\alpha=\delta^{-1} \|\Phi(\bfu')/R-\bfx\|$ for the lower-dimensional feature space, $\delta=1/3$.}\label{tab:accuracy_lower}
\end{table}

\begin{table}
	\centering
	\small
	\begin{tabular}{|c|c|c|c|c|c|c|c|c|c|c|}
		\hline
		$\#$ & 1 & 2 & 3 & 4 & 5 & 6 & 7 & 8 & 9 & 10 \\
		\hline\hline
		$\alpha$    & 0.090 & 0.205 & 0.229 & 0.032 & 0.097 &  0.046 &  0.106 & 0.053 & 0.193 & 0.139 \\
		$n$    & 33 & 39 & 34 & 35 & 37 &  36 & 34 & 36 & 36 & 35 \\
		\hline
	\end{tabular} 
	
	\vspace{5mm}
	
	\begin{tabular}{|c|c|c|c|c|c|c|c|c|c|c|}
		\hline
		$\#$ & 11 & 12 & 13 & 14 & 15 & 16 & 17 & 18 & 19 & 20 \\
		\hline\hline
		$\alpha$    & 0.063 & 0.001 & 0.051 &  0.305 &  0.085 &  0.084 & 0.065 & 0.169 & 0.121 & 0.140 \\
		$n$    & 31 & 30 & 37 & 32 & 33 &  33 & 36 & 34 & 38 & 33\\
		\hline
	\end{tabular} 
	\vspace{5mm}
	
	\caption{Accuracy $\alpha$ of finding triggers expressed as $\alpha=\delta^{-1} \|\Phi(\bfu')/R-\bfx\|$ for the lower-dimensional feature space, $\delta=2/3$.}\label{tab:accuracy_lower_2}
\end{table}

Empirical values of the accuracy of implementation of  these attacks, $\alpha$, for lower-dimensional feature maps produced by mappings (\ref{eq:reduced_dimension}) are shown in Tables \ref{tab:accuracy_lower}, \ref{tab:accuracy_lower_2}.

These figures, along with the fact that all these attacks returned a successful outcome (for Scenario $1$), enable us to illustrate how conservative the bounds provided in Theorem \ref{thm:plain} could be. 

Indeed, computing the term
\[
P_1(\alpha,\delta,\gamma,n)=\pi^{-1/2} \frac{\Gamma\left(\frac{n}{2}\right)}{\Gamma\left(\frac{n-1}{2}\right)} \int_{0}^{\arccos{(\varphi(\gamma,\delta,\alpha) )}} \sin^{n-2}(\theta)d\theta
\]
for $\gamma=0.9$ in bound (\ref{eq:probability_plain}) for
$\alpha,n$ taken from the first experiment in Table \ref{tab:accuracy} (column corresponding to $\#=1$ in Table \ref{tab:accuracy}) returns $P_1(\alpha,\delta,\gamma,n)=0.1561$. Substituting this value into (\ref{eq:probability_plain}) for $M>10$ produces a negative number rendering the bound not useful in this case. The same comment applies to other experiments. At the same time, as our experiments confirm, the attacks turn out to be successful in practice. If, however, we set $\alpha=0$ then the value of $P_1(0,\delta,\gamma,n)$ for the first experiment becomes $4.9180\times 10^{-4}$.

These observations highlight limitations of the bounds in Theorems~\ref{thm:plain} and \ref{thm:incremental} for determining the  probability of success in practice when $\alpha$ is large. They also show the additional relevance of 
Theorem~\ref{thm:concentration_collapse} and the ``concentrational collapse'' effect for developing better understanding of conditions leading to increased vulnerabilities to stealth attacks (see Tables \ref{tab:collapse_illustration} and \ref{tab:collapse_illustration_reduced} illustrating the difference between bounds in Theorems~\ref{thm:plain}, \ref{thm:incremental} and  \ref{thm:concentration_collapse}). 

\subsection{Selection of a neuron to attack}\label{sec:neuron_selection}

Let $L$ be the layer in which a neuron is to be selected for the attack. We suppose that the network has a layer $L+1$ in its computational graph. Suppose that $w^{[L]}_{i,j}$ denotes the $j$th weight of the $i$th neuron in layer $L$, and the total number of neurons in layer $L$ is $N_L$.

We say that the {\it output} weight vector of the $i$-th neuron in layer $L$ is a vector
\[
\bar{\bfw}_i^{[L]}=\left(\begin{array}{c}
	w^{[L+1]}_{1,i}\\
	w^{[L+1]}_{2,i}\\
	\vdots \\
	w^{[L+1]}_{N_L,i}
\end{array}\right).
\]

{\it Determining neuron rank.} For each $i=1,\dots,N_L$, we computed $L_1$-norms of $\bar{\bfw}_i^{[L]}$:
\[
\|\bar{\bfw}_i^{[L]}\|_1=\sum_{j=1}^{N_{L+1}} |w_{j,i}^{L+1}|
\]
and created a list  of indices $n_1,n_2,\dots,n_{N_L}$ such that
\[
\|\bar{\bfw}_{n_1}^{[L]}\|_1 \leq \|\bar{\bfw}_{n_2}^{[L]}\|_1 \leq \cdots \leq \|\bar{\bfw}_{n_{N_L}}^{[L]}\|_1.
\]

The susceptibility rank of the $i$-th neuron, $\mbox{Rank}(i)$, is defined as a number $k\in\{1,\dots,N_L\}$:
\[
\mbox{Rank}(i)=k: \ i=n_k.
\]
In our code, any ties were broken at random.

\end{document}